\documentclass[twocolumngrid,prd,superscriptsize,reprint]{revtex4-1}
\usepackage[utf8]{inputenc} 
\usepackage{graphicx,xcolor,overpic,mathtools}
\usepackage{amsthm,amsmath,amssymb}
\usepackage[colorlinks]{hyperref}
\usepackage{textcomp}
\definecolor{coolblack}{rgb}{0.0, 0.18, 0.39}
\hypersetup{
    colorlinks = false,
   }
\PassOptionsToPackage{normalem}{ulem}
\usepackage{ulem} 
\usepackage{sidenotes}
\usepackage{bm}
\usepackage{url}
\usepackage{physics}
\usepackage[makeroom]{cancel}
\usepackage{cleveref}
\usepackage{tensor}
\usepackage{mathrsfs}
\usepackage{slashed}
\usepackage{caption}
\usepackage{subcaption}
\usepackage[mode=buildnew]{standalone}

\newcommand{\comment}[1]{}

\usepackage{pgfplots}

\usepackage{diagbox}
\usepackage{stackengine,amssymb,graphicx}

\NewDocumentCommand{\evat}{sO{\bigg}mm}{%
  \IfBooleanTF{#1}
   {\mleft. #3 \mright|_{#4}}
   {#3#2|_{#4}}%
}
\usepackage{enumerate}
\definecolor{azure}{rgb}{0.0, 0.5, 1.0}

\begin{document}

\title[]{Effective no-hair relations for spinning Boson Stars}

\author{Christoph Adam}
\author{Jorge Castelo}
\author{Alberto Garc\'ia Mart\'in-Caro}
\author{Miguel Huidobro}%
\affiliation{%
Departamento de F\'isica de Part\'iculas, Universidad de Santiago de Compostela and Instituto
Galego de F\'isica de Altas Enerxias (IGFAE) E-15782 Santiago de Compostela, Spain
}%
\author{Andrzej Wereszczynski}
\affiliation{
Institute of Physics, Jagiellonian University, Lojasiewicza 11, Krak\'ow, Poland
}%

\date[ Date: ]{\today}
\begin{abstract}

Boson Stars are, at present, hypothetical compact stellar objects whose existence, however, could resolve several enigmas of current astrophysics. If they exist, either as independent astrophysical entities or as a matter admixture of more standard compact stars, then their imprints can probably be observed in the not-too-distant future from the gravitational signal of coalescing binaries in current and future GW detectors.
Here we show that the multipole moments of rotating boson stars obey certain universal relations, valid for a broad set of models and various states in terms of the \textit{harmonic indices}.
These universal relations are equivalent to a kind of no-hair theorem for this exotic matter, 
allowing to map these universal (i.e. model independent) multipoles to an equally universal gravitational field around the stellar object.
Further, the multipole moments can be related to observable astrophysical quantities.

\vspace{0.5cm}
\large{\bf Dedicated to the memory of our unforgettable friend and colleague Ricardo Vázquez.}  

\end{abstract}

\maketitle

\begin{quote}
 
\end{quote}

\tableofcontents

\section{Introduction}


The modeling of particles or other, more macroscopic compact matter objects by smooth, stable lumps of field energy bound together by their self-interaction is a very attractive theoretical idea, which has stimulated intense research activities over many decades. In the particular case of {\em scalar} fields where at least part of the binding is provided by gravitation, the resulting objects are known as Boson Stars (BS).
The (electromagnetic) "geons" of Wheeler \cite{PhysRev.97.511} probably constitute the first case of gravitationally bound field lumps, and also the investigation of proper BS started already more than five decades ago \cite{PhysRev.172.1331,PhysRev.187.1767,PhysRev.148.1269} and grew steadily ever since.
More recently, the discovery of the Higgs boson at CERN \cite{ATLAS:2012yve,CMS:2012qbp} demonstrated that fundamental scalar fields are not only interesting theoretically, but do exist in nature.
As a result, the interest in the topic of BS has increased even further. 

BS arise in simple field theoretical models like massive complex bosonic fields, both for scalars \cite{PhysRev.172.1331,PhysRev.187.1767,PhysRev.148.1269} and vectors  (Proca stars) \cite{Brito:2015pxa}. 
Both the phenomenology and the properties of these compact objects (for reviews see \cite{Liebling:2012fv,Lai:2004fw,Schunck:2003kk}) and the dynamical mechanisms by which the BS are formed \cite{Jetzer:1990xa,Jetzer:1991jr,Seidel:1993zk,DiGiovanni:2018bvo,Sanchis-Gual:2019ljs} and their stability \cite{Brito:2015pxa,GLEISER1989733,Sanchis-Gual:2017bhw,Khlopov:1985jw, Sanchis-Gual:2021phr,Siemonsen:2020hcg} have been extensively studied.
BS properties strongly depend on the Lagrangian, and various types of potentials encode different self-interactions among the fields, allowing to model several astrophysical systems, from Neutron-Star like objects to dark matter galaxy haloes \cite{Schunck:1998nq}, without forgetting their prominent role as Black Hole (BH) mimickers  \cite{PhysRevD.80.084023,Herdeiro:2021lwl} and intermediate-mass astrophysical objects.
Interest in BS is also related to the possible existence of dark-matter ultralight scalar bosons \cite{Freitas:2021cfi},  or extensions of the Standard Model such as the axion \cite{PhysRevLett.40.223,PhysRevLett.40.279}. 

The formation of astrophysical objects is almost always accompanied by a nonzero angular momentum and, therefore, rotation is fundamental from an astrophysical point of view. The resulting more general and realistic models of rotating BS have been developed and studied both for the scalar \cite{Schunck:1996he,Yoshida:1997qf} and vectorial cases \cite{Brito:2015pxa,Herdeiro:2016tmi}. These axisymmetric spinning BS (SBS) have also been studied from a phenomenological point of view \cite{Vincent:2015xta}, and their stability and dynamical properties are explored in \cite{Sanchis-Gual:2019ljs,Sanchis-Gual:2021phr}.
Other, more exotic, generalizations take into account, e.g., generalized models of gravity like Palatini gravity \cite{Maso-Ferrando:2021ngp}, Einstein-Gauss-Bonnet theory, scalar-tensor models \cite{PhysRevD.56.3478}, or the semiclassical gravity framework \cite{Alcubierre:2022rgp}. Some more exotic BS models like multi-state boson stars \cite{Urena-Lopez:2010zva}, $\ell$-Boson stars \cite{Alcubierre:2018ahf}, Scalaroca stars \cite{Pombo:2023xkw} or even the Proca-Higgs Stars \cite{Herdeiro:2023lze} have been considered in the last years, among other cases.

Since the LIGO-VIRGO collaboration reported the first event \cite{LIGOScientific:2016aoc}, gravitational wave (GW) astronomy has become one of the most powerful tools for the study of the Universe. Nowadays, advanced LIGO and Virgo, or KAGRA have reported a multitude of events \cite{LIGOScientific:2021hvc,KAGRA:2018plz}, involving binary BH, binary neutron stars (NS) \cite{LIGOScientific:2018cki}, NS-BH mergers, and even events where the merging objects are not well identified yet. One of the last mentioned events, measured in 2020 by advanced LIGO-VIRGO, could be potentially explained as a head-on collision of two Proca stars \cite{Bustillo:2020syj}. The dynamical situation where two BS are orbiting each other has been studied with the aim of extracting the waveforms \cite{Palenzuela:2007dm,Sanchis-Gual:2018oui},
and the merger scenario is nowadays a vibrant field of research \cite{Palenzuela:2006wp,Palenzuela:2007dm,Sanchis-Gual:2018oui,Bezares:2022obu}. The possibility that BS are compact astrophysical sources
different from BH and NS is one of the main lines of research in the GW community \cite{Cardoso:2019rvt,LIGOScientific:2021sio,Maggio:2021ans,CalderonBustillo:2020fyi}. Other kinds of emissions, like  axion electric emission \cite{Sanchis-Gual_axion_electric}, are also studied when considering some very specific models. 

For the case of NS, in the last decade certain universal relations have been established, i.e., relations between different observables which do not depend on the particular equation of state (EOS) used for the description of the NS. The most famous of these relations is the so-called $I$-Love-$Q$ relation, proposed by Yagi and Yunes in \cite{Yagi:2013awa}, which relates the moment of inertia $I$, the tidal deformability (Love number) \cite{Hinderer:2007mb,Postnikov:2010yn} 
and the quadrupolar moment $Q$. These relations are important for several reasons. First of all, if their validity is assumed, they allow 
to extract observable quantities that are difficult to measure.
Further, they are useful for the breaking of certain degeneracies between the NS spin parameter and the quadrupolar moment in binary systems \cite{Yagi:2013awa,Yagi:2016bkt}.


These relations have been completed \cite{Reina:2017mbi}, well tested \cite{Adam:2020aza}  and extended to high spin and magnetic NS \cite{Haskell:2013vha}, and also to modified gravity theories
\cite{Sham:2013cya,Chakravarti:2019aup,Doneva:2017jop}. Other quasi-universal relations, involving higher multipoles and Love numbers \cite{Yagi:2013sva,Godzieba:2021vnz}, the compactness, gravitational binding energies \cite{UnivRelsbinding}, and oscillation frequencies of (quasi) normal modes \cite{Torres-Forne:2019zwz} have been studied \cite{Sun:2020qkj,Doneva:2017jop}. For the case of BS, some first universal relations have been shown recently in \cite{Adam:2022nlq,Vaglio:2022flq}.

 An observational confirmation of these 
 relations is quite challenging because of the technical difficulties when measuring the involved properties even for NS.
 Still, this field of observational astrophysics has accumulated more than forty years of development, and there exist many promising proposals for the measurement of the spins, moments of inertia \cite{Kramer:2009zza,Silva:2020acr,Silva:2016myw,Link:1999ca,Andersson2012PulsarGT,Chamel2013CrustalEA,Steiner:2014pda,Damour:1988mr,Lattimer:2004nj,Bejger:2005jy}, multipole moments and sizes   \cite{Zhao:2022tcw,abbott2017gw170817,Silva:2020acr}
 using NICER and GW data.

 In addition to their astrophysical relevance, the mere existence of universal relations is essential from a purely theoretical point of view. The understanding of the universal behavior as an effective no-hair theorem is transcendental in what follows. The no-hair theorems \cite{misner1973gravitation,Robinson:1975bv,Israel:1967wq,Hawking:1971tu,hawking1972black,Carter:1971zc} formulate that stationary axisymmetric BHs are fully described by  their mass, spin, angular momentum, and charge. A Kerr BH exterior gravitational field can be  reconstructed as an infinite series of multipoles, depending only on the mass-monopole and the current-dipole \cite{Geroch:1970cd,Hansen:1974zz}. The importance of the multipoles lies not only in their link with the gravitational field created by an object but also  in their direct relation with astrophysical observables \cite{Ryan2,Ryan3,Pappas:2012nt}. For NS and Quark Stars, the BH no-hair theorems do not apply, as they are non-vacuum sources, but universal and quasi-universal relations do. The way of understanding them like effective no-hair theorems for fermionic compact objects was fully treated in several works \cite{Yagi:2014bxa,Doneva:2017jop,Stein:2013ofa,Yagi:2013awa,Yagi:2016bkt}. A similar treatment for BS for the first two multipoles was presented in \cite{Adam:2022nlq,Vaglio:2022flq}.

The present paper extends and generalizes the results of  \cite{Adam:2022nlq}, investigating the existence of approximately model-independent, effective no-hair relations among the multipole moments up to the hexadecapole order for rapidly rotating BS. Further, we analyze BS for the first three and most representative harmonic indices.

 To do so, we solve the Einstein equations for complex scalar rotating BS with the  FIDISOL-CADSOL code \cite{fidisol,schonauer1989efficient,schonauer2001we}. As BS are infinitely extended objects without any particular surface \cite{Liebling:2012fv}, we identify radii with the perimetral radius $R_{99}$ that contains $99\%$ of the BS mass $M_{99}$ \cite{Delgado:2020udb}.  We use units where $\hbar=c=1$.

The structure of the paper is the following. In  \cref{setup} we introduce
the theoretical set-up. We present the numerical scheme in  \cref{numerical}. In \cref{multi} we show how to obtain the multipolar expansion for our stationary and axisymmetric space-time systems and other observables of interest like the moment of inertia $I$. In  \cref{results} we present our results concerning the discovered universal relations and compare with NS results. Finally,  \cref{conclusions} contains our conclusions.


\section{Theoretical set-up}\label{setup}
As in our previous work \cite{Adam:2022nlq}, the system is described by the Einstein-Klein-Gordon (EKG) action, where a  massive complex scalar field $\Phi$ is minimally coupled to the Einstein gravity \cite{Liebling:2012fv},
\begin{equation}
    \mathcal{
    S}=\int \left(\frac{1}{16\pi G}R+\mathcal{L}_{\Phi}\right)\sqrt{-g}d^4x.
    \label{action}
\end{equation}
Here $g$ is the metric determinant, $R$ the Ricci scalar, and the Lagrangian  that governs the field dynamics reads,
\begin{equation}
 \mathcal{L}_{\Phi}=-\frac{1}{2}\left[g^{\alpha\beta}\nabla_{\alpha}\Phi^*\nabla_{\beta}\Phi+V\left(|\Phi|^2\right)\right].
    \label{lagrangian}
\end{equation}
The potential $V\left(|\Phi|^2\right)$ depends only on the absolute value of the scalar field, and respects the global $U(1)$ invariance of the model. All potentials we consider contain the quadratic mass term $\mu^2|\Phi|^2$, and various self-interactions. The scalar potential for the BS plays an analogous role to the EOS in the NS case. In this work, we use the same models as in \cite{Adam:2022nlq}, but we expand the range of the coupling constants for some of them, allowing to reach higher masses and second  branches of solutions in some cases. Despite the use of some higher values for the quartic self-interaction constant, we leave the complete analysis of that regime for future work. All details about the different potentials are shown in \cref{appendixA}. 

By varying the action (\ref{action}), we find the the EKG equations
\begin{equation}
\begin{split}
    &R_{\alpha\beta}-\frac{1}{2}Rg_{\alpha\beta}=8\pi T_{\alpha\beta}, \\
    &
    g^{\alpha\beta}\nabla_\alpha\nabla_{\beta}\Phi=\frac{dV}{d|\Phi|^2}\Phi,
\label{kg}
\end{split}
\end{equation}
where $R_{\alpha\beta}$ is the Ricci tensor and $T_{\alpha\beta}$ is the canonical Stress-Energy tensor of the scalar field,
\begin{equation}
    T_{\alpha\beta}=2\nabla_{(\alpha}\Phi^*\nabla_{\beta)}\Phi-2g_{\alpha\beta}\left[g^{\mu\nu}\nabla_{(\mu}\Phi^*\nabla_{\nu)}\Phi+V\left(|\Phi|^2\right)\right]].
    \label{stress}
\end{equation}
For the above Stress-Energy tensor to satisfy stationarity and axial symmetry, the scalar field ansatz takes the form
\begin{equation}
     \Phi(t,r,\theta,\psi)=\phi(r,\theta)e^{-i(w t+n\psi)}
     \label{scalar}
 \end{equation}
where $w \in \mathbb{R}$ is the angular frequency of the field, and $n \in \mathbb{Z}$ (also called $m$ or  $s$ in the literature \cite{Vaglio:2022flq,Ryan:1996nk}) is the \textit{azimutal harmonic index}, also called \textit{azimutal winding number}. This parameter enters the problem as an integer related to the star's angular momentum. Finally,  $\phi(r,\theta)$ is the profile of the star.
We assume the following ansatz for the metric, describing the stationary 
and axisymmetric space-time\cite{Herdeiro:2015gia,PhysRevD.55.6081},
\begin{equation}
\begin{split}
    ds^2=&-e^{2\nu}dt^2+e^{2\beta}r^2\sin^2\theta\left(d\psi-\frac{W}{r}dt\right)^2\\
    &+e^{2\alpha}(dr^2+r^2d\theta^2).
    \label{Herdeiro}
    \end{split}
\end{equation}
Here, $\nu, \alpha, \beta$ and $W$ are functions which depend only on $r,\theta$. 
 
The universal I-love-Q relation has been discovered in the context of NS. To verify its existence for BS, one has to underline two crucial differences. 
Firstly, BS are derived in a full field theoretical framework, where the matter forming BS follows its field equation. On the contrary, NS are typically obtained by assuming a given equation of state (stress-energy tensor) describing nuclear matter at a specific range of densities. However, field theoretical models have also been proposed. 
Secondly, BS do not allow for a smooth transition from rapid to slow rotation, at least using our ansatz \cite{Lai:2004fw} due to the discretized angular momentum, which, together with discrete on-axis regularity conditions, do not allow for a perturbative transition as Kobayashi, Kasai, and Futamase first proved \cite{Kobayashi:1994qi}. On the other hand, such a limit is the main ingredient of the Hartle-Thorne formalism \cite{Hartle:1967he,Hartle:1968si}, which provides the most straightforward approach for deriving quadrupole moment $Q$ and love numbers for rotating NS, and therefore, to establish the I-Love-Q relation for slowly rotating NS. In our approach, we have to work in a full rotation context, where we first solve the system and then, as we will see in \cref{multi}, we obtain the multipolar expansion.

\section{Numerical implementation }\label{numerical}

To perform the numerical integration of the EKG system, we first rescale the radial distance and angular frequency by the mass $\mu$ of the boson field, $
r\rightarrow r\mu, \hspace{0.2cm}w\rightarrow w/\mu$. This redefinition of the length removes the explicit $\mu$ dependence from the field equations but changes the coupling constant definitions for the different potentials.
We also rescale the field $ \phi\rightarrow\phi\sqrt{4\pi}$ for simplicity. 

The mathematical problem we have to solve is a set of five coupled, non-linear partial differential equations for the metric functions and the scalar field, which follows from \cref{kg}. We also take into account the constraints, $E^r_{\theta}=0, E^r_r-E^{\theta}_{\theta}=0$, where $E^{\mu}_{\nu}=R^{\mu}_{\nu}-\frac{1}{2}Rg^{\mu}_{\nu}-2 T^{\mu}_{\nu}$. To perform the numerical integration, we use the FIDISOL/CADSOL package \cite{fidisol,schonauer1989efficient,schonauer2001we}, a Newton-Raphson-based code with an arbitrary grid and
consistency order. It also  provides an error estimate for each unknown function. 
We use the EKG system in the following form:
\begin{equation}
\begin{split}
&-e^{2\alpha}\frac{r^2}{2}\sin^2(\theta)\left(-E_{t}^{t}+E_{r}^{r}+E_{\theta}^{\theta}-E_{\phi}^{\phi}\right)=0\\
&e^{2\alpha}\frac{r^2}{2}\sin^2(\theta)\left(E_{t}^{t}+E_{r}^{r}+E_{\theta}^{\theta}-E_{\phi}^{\phi}+\frac{2WE_{\phi}^{t}}{r}\right)=0\\
&e^{2\alpha}\frac{r^2}{2}\sin^2(\theta)\left(-E_{t}^{t}+E_{r}^{r}+E_{\theta}^{\theta}-E_{\phi}^{\phi}-\frac{2WE_{\phi}^{t}}{r}\right)=0\\
&2re^{2\nu +2\alpha-2\beta}E_{\phi}^{t}=0\\
& \frac{e^{2\alpha}r^2\sin^2(\theta)}{\phi} \Phi^*\left(\Box-\frac{d V}{d|\phi|^2} \right) \Phi=0.
\end{split}
\label{EKG-system}
\end{equation}
   
\begin{figure*}[]
\subfloat{%
  \hspace*{1.0cm}\includegraphics[clip,width=1.0\columnwidth]{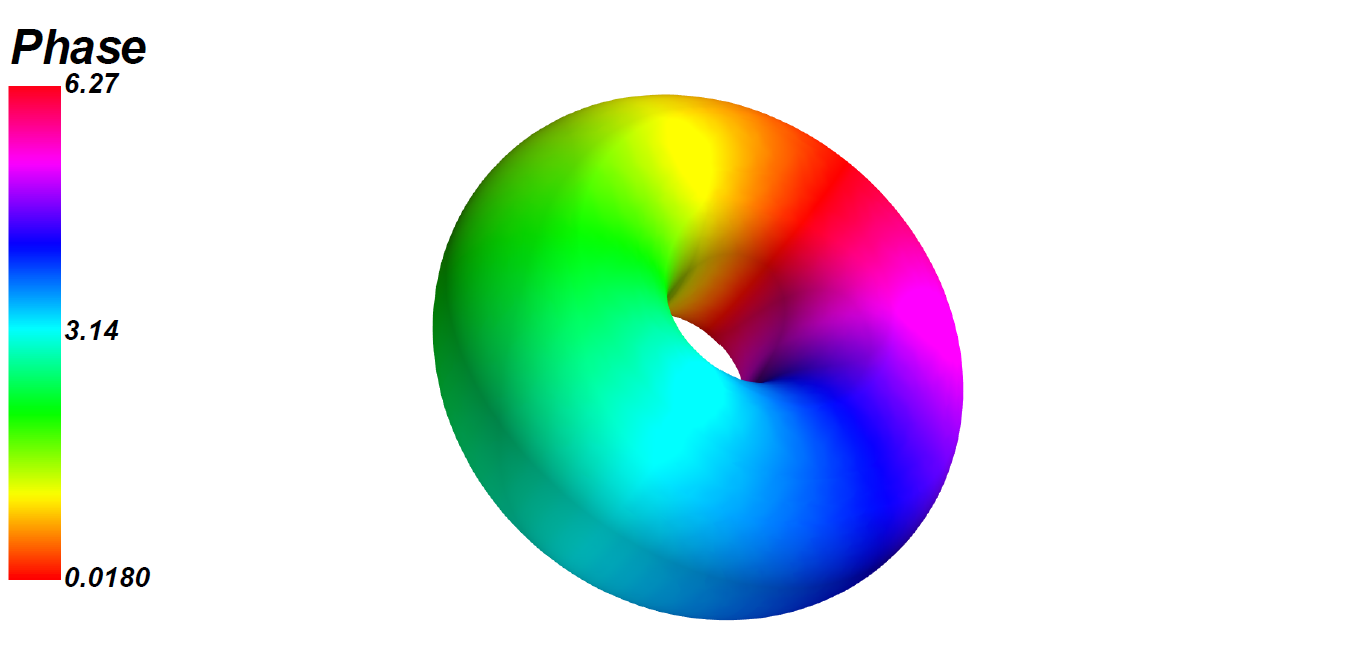}%
}

\subfloat{%
  \hspace*{1.0cm}\includegraphics[clip,width=1.0\columnwidth]{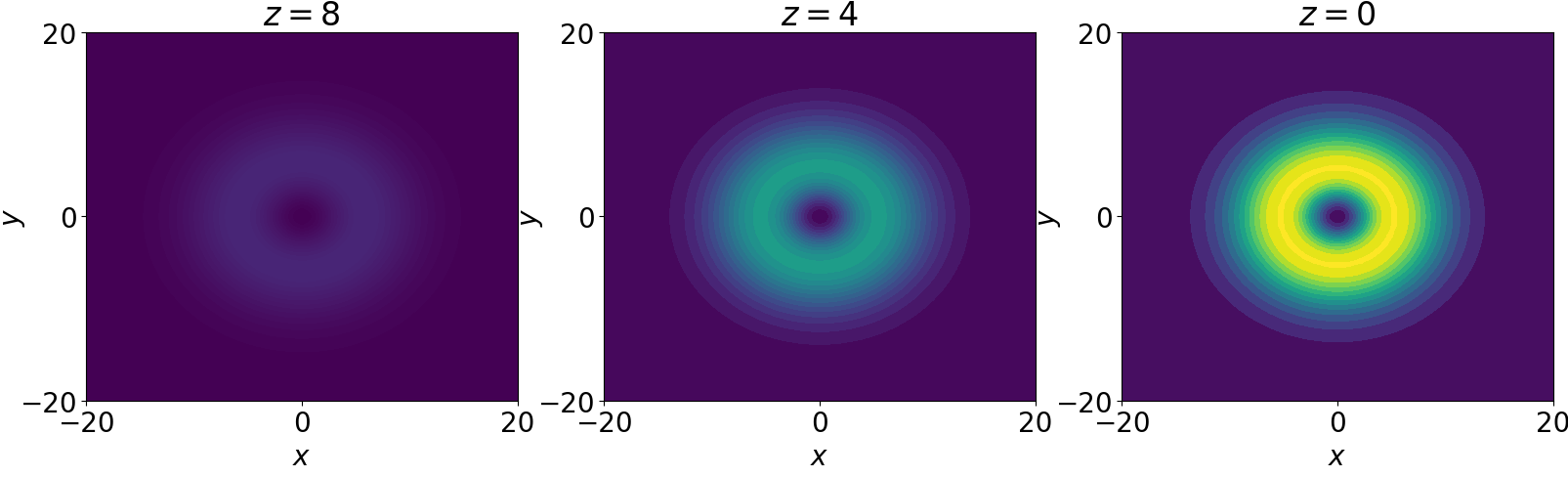}%
}
\caption{SBS for a quartic interaction potential with $\lambda=50$. The field density is shown in the 3D space in the upper plot. Different colors represent the field phase $\psi$. In the lower plot, for the same star, we show the field density in three different layers with respect to the axis of symmetry. Each layer was taken at a different distance from the equator, being the first one the farthest, then we have a second slice closer to the equator, and the third slice cuts through the equatorial plane, where the 
 field density is highest. }
\label{bSs}
\end{figure*}

We compactify the radial coordinate by the following definition $x\equiv r/(1+r)$ moving from $r\in [0,\infty)$ to a finite segment $x \in [0,1]$. After discretizing the equations on a $(401\times 40)$, $(x,\theta)$ grid, where $0\leq x\leq 1$ and $0\leq \theta\leq \pi/2$, we impose boundary conditions on the field profile and the metric functions. Asymptotic flatness reads,
\begin{equation}
    \lim_{r\rightarrow\infty} \alpha =\lim_{r\rightarrow\infty} \beta =\lim_{r\rightarrow\infty} \nu =\lim_{r\rightarrow\infty} W =\lim_{r\rightarrow\infty} \phi = 0.
\end{equation}
Reflection on the rotation axis and  axial symmetry implies that at $\theta=0$ and $\theta=\pi$,
\begin{eqnarray}
    \partial_{\theta}\alpha=\partial_{\theta}\beta=\partial_{\theta}\nu=\partial_{\theta}W=\partial_{\theta}\phi=0. 
\end{eqnarray}
Since the solutions have to be symmetric with respect to a reflection along the equatorial plane, this condition is also obeyed on the equatorial plane, $\theta=\pi/2$.
Eventually, regularity at the origin requires $\partial_r \alpha=\partial_r \beta=\partial_r \nu =W= \phi=0$ when $r \to 0$, and regularity in the symmetry axis further imposes $\left.  \alpha=\beta \right|_{\theta=0, \pi}$ \cite{Herdeiro:2015gia}.
Further details about the solver are explained in  \cite{Delgado:2022pwo,Adam:2022nlq}. After performing the numerical integration of the EKG equations, we can appreciate how the BS presents a toroidal field distribution. In \cref{bSs} we show a solution for the $w=0.9, n=1$ star, for the quartic self-interaction potential with $\lambda=50$.

\begin{figure*}[]
\centering
\hspace*{-0.0cm}\includegraphics[width=0.50\textwidth]{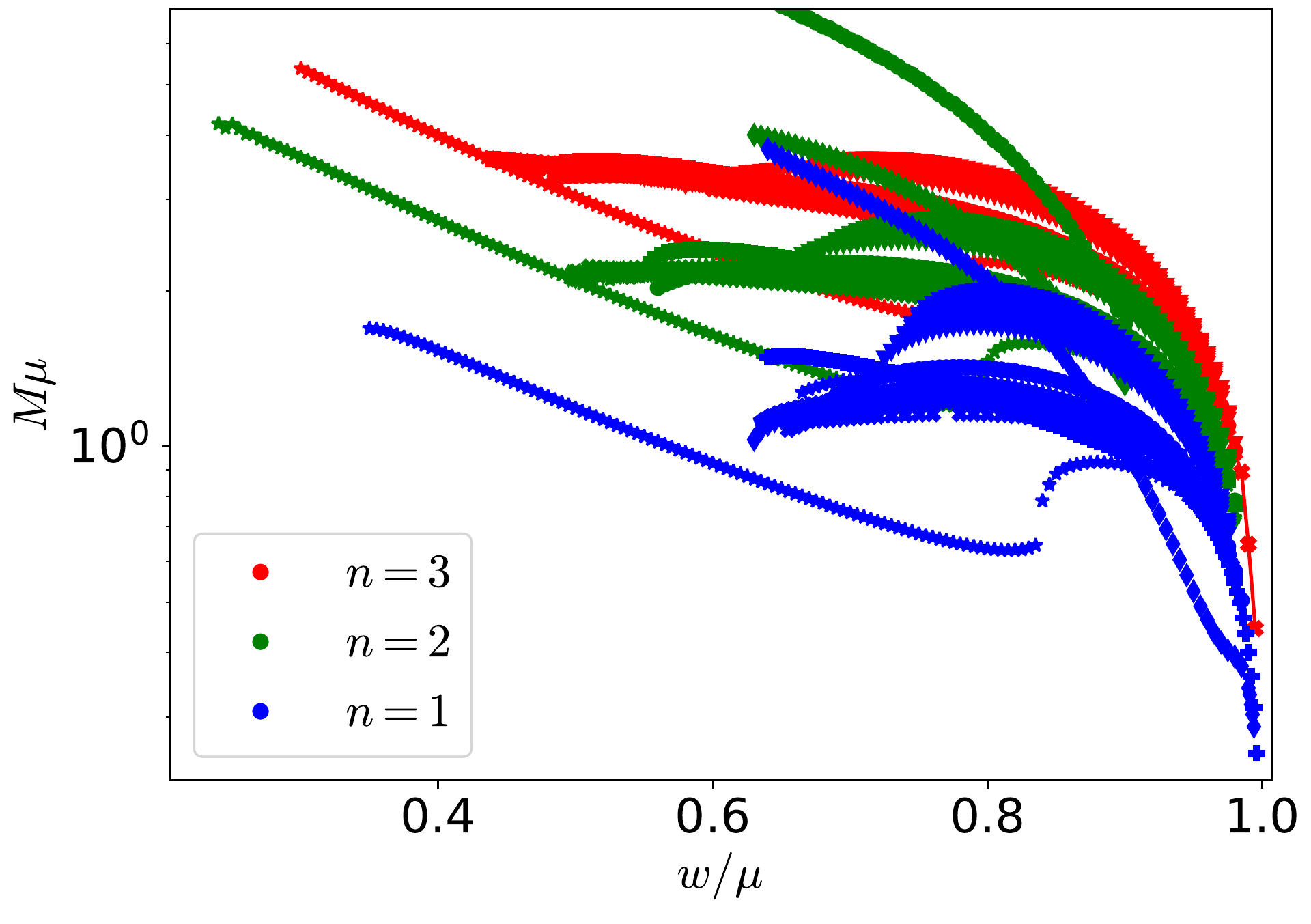}
\caption{Mass frequency curves for the different models we have employed. Also, for different harmonic indexes. The legend shows that red dots are $n=3$ stars, greens are $n=2$, and blues are $n=1$. The different potentials for obtaining those representative curves are shown in the appendix. As can be appreciated from the plot, we have some secondary branches and also very different curves, making our set very wide-ranging.}
\label{mbS}
\end{figure*}

We have used considerable data, using all the potentials in \cref{appendixA}. The resulting Mass vs. field frequency plots are shown in \cref{mbS} in agreement with \cite{Siemonsen:2020hcg,Delgado:2020udb,Herdeiro:2015gia}. As can be seen, we find stars ranging from $\sim 0.2 M\mu$ in the limit $w\rightarrow 1$ to very massive stars with $\sim 6 M\mu$ for some potentials. This implies that for values of $\mu$ ranging from $10^{-9} \mathrm{eV}$ to $10^{-12} \mathrm{eV}$ we could model objects with masses between $0.001 M_{\odot}$ and $ 10  M_{\odot}$. An in-depth study of the most massive BS models was developed in \cite{Ontanon:2021hbg}.

\section{Multipolar structure and global properties.}\label{multi}

In General Relativity, multipoles result from two sources, the energy density and the current density \cite{RevModPhys.52.299,PhysRevD.52.821}. They play a fundamental role both from the theoretical and the astrophysical point of view. 
The basics of the metric-multipole-expansion were developed in
\cite{Geroch:1970cd,Hansen:1974zz,fodor1989multipole}, where the last approach was used in the NS framework in \cite{Pappas:2018csu,1976ApJ...204..200B}. We follow this approach,   but adapted to our bosonic scenario.
\subsection{Multipole moments}

We reparametrize our metric functions as $\omega=\frac{W}{r},\hspace{0.4cm}B=e^{\nu+\beta}$.
The following expressions provide a consistent asymptotic multipolar expansion of the metric functions (see \cite{1976ApJ...204..200B,morse1954methods}),
\begin{eqnarray}
        \nu&=&\sum_{l=0}^{\infty}\bar{\nu}_{2l}(r)P_{2l}(\cos\theta),\hspace{0.4cm}   \hspace{0.9cm} \bar{\nu}_{2l}(r)= \sum_{k=0}^{\infty}\frac{\nu_{2l,k}}{r^{2l+1+k}}, \nonumber  \\
         \omega&=&\sum_{l=0}^{\infty}\bar{\omega}_{2l-1}(r)\frac{dP_{2l-1}(\cos\theta)}{d\cos\theta},     \hspace{0.2cm} \bar{\omega}_{2l-1}(r)= \sum_{k=0}^{\infty}\frac{\omega_{2l-1,k}}{r^{2l+1+k}} \nonumber  \\
          B&=&1+\sum_{l=0}^{\infty}\bar{B}_{2l}(r)T_{2l}^{\frac{1}{2}}(\cos\theta),     \hspace{0.2cm}     \hspace{0.2cm} \bar{B}_{2l}(r)= \frac{B_{2l}}{r^{2l+2}},       
    \label{multipoles}
\end{eqnarray}
where $P_l(\cos\theta)$ and $T_l^{\frac{1}{2}}(\cos\theta)$ are the Legendre and Gegenbauer polynomials, respectively. 
As these coefficients are crucial in our analysis, let us explain how we obtain them. Instead of the full source integration (see, e.g.,  \cite{PhysRevD.55.6081,Doneva:2017jop}), we use the fact that we already solved the EKG system numerically and, hence, know the functions $\nu$, $\omega$ and $B$. The multipole coefficients are then found by integrating over the angles after projecting on the appropriate polynomial and taking the corresponding radial limits.

Indeed, $\nu_{2l,0}=\lim_{r\rightarrow\infty}\left(r^{2l+1}\nu_{2l}\right)$, where $\nu_{2l}$ can be found by an appropriate projection on the Legendre polynomial. Hence, 
\begin{equation}
\nu_{2l,0}=N_{2l}\lim_{r\rightarrow\infty}r^{2l+1}\int_{-1}^{1}\nu(r,\theta)P_{2l}(\cos\theta)d\cos\theta,
    \label{coeffint}
\end{equation}
where $N_{2l}$ is the normalization constant. Following the same procedure, we obtain the expressions for $B_{2l}$ and $\omega_{2l-1,0}$. We explicitly find, to order $l=2$,
\begin{equation}
\begin{split}
&
\nu_{0,0}=\frac{1}{2}\lim_{r\rightarrow\infty}r\int_{-1}^{1}\nu(r,\theta)dy,\\
&
\nu_{2,0}=\frac{5}{2}\lim_{r\rightarrow\infty}r^{3}\int_{-1}^{1}\nu(r,\theta)\frac{(3y^2-1)}{2}dy,\\
&
\nu_{4,0}=\frac{9}{2}\lim_{r\rightarrow\infty}r^{5}\int_{-1}^{1}\nu(r,\theta)\frac{1}{8}(35y^4-30y^2+3)dy,\\
&
\omega_{1,0}=\frac{3}{4}\lim_{r\rightarrow\infty}r^{3}\int_{-1}^{1}\omega(r,\theta)(1-y^2)dy,\\
&
\omega_{3,0}=\frac{7}{24}\lim_{r\rightarrow\infty}r^{5}\int_{-1}^{1}\omega(r,\theta)(1-y^2)\frac{3}{2}(5y^2-1)dy,\\
&
B_{0}=\lim_{r\rightarrow\infty}r^{2}\int_{-1}^{1}(B(r,\theta)-1)\sin\theta\sqrt{\frac{2}{\pi}}dy,\\
&
B_{2}=\lim_{r\rightarrow\infty}r^{4}\int_{-1}^{1}(B(r,\theta)-1)\sin\theta\sqrt{\frac{2}{\pi}}(4y^2-1)dy,
\end{split}
\label{coeffint2}
\end{equation}
where we have redefined $y=\cos\theta$. We show the orthogonality relations between the polynomials and a list with a few of them in \cref{appendixB}.
 Now, the multipoles can be obtained as combinations of the expansion coefficients in (\ref{multipoles}), see \cite{Pappas:2014gca} for details. Specifically, one can show that the multipoles up to $l=2$ are, 
\begin{equation}
\begin{split}
 &M_0=M = -\nu_{0,0},\\
 &S_1=J =\frac{\omega_{1,0}}{2},\\
 &M_2=Q=\frac{4}{3}B_{0}\nu_{0,0}+\frac{\nu_{0,0}^3}{3}-\nu_{2,0},\\
 &S_3=-\frac{6}{5}B_0\omega_{1,0}-\frac{3}{10}\nu^2_{0,0}\omega_{1,0}+\frac{3}{2}\omega_{3,0},\\
 &M_4 =-\frac{32}{21}B_0\nu^3_{0,0}-\frac{16}{5}B^2_0\nu_{0,0}+\frac{64}{35}B_2\nu_{0,0}+\frac{24}{7}B_0\nu_{2,0}\\
   &+\frac{3}{70}\nu_{0,0}\omega^2_{1,0}-\frac{19}{105}\nu^5_{0,0}+\frac{8}{7}\nu_{2,0}\nu^2_{0,0}-\nu_{4,0}.
    \label{M1}
    \end{split}
\end{equation}

\subsection{Numerical tools for the calculation}
In comparison to our previous work \cite{Adam:2022nlq}, we have developed an improved technique when extracting the different coefficients from the numerical simulations. This fact is reflected in our new fittings and allows us to have a better accuracy when obtaining the $I,\chi,Q$, making it also possible to obtain $S_3$ and $M_4$. The new calculations were done with a bigger number of grid points for each star. But we also have to take into account that for each model the size of the star varies, so the limit when $r\rightarrow \infty$ must be taken carefully, looking for an equilibrium between the distance to the matter region and the number of points in order to have a good fit. We also found that the fittings for the higher coefficients in the metric expansions, i.e. $\nu_{4,0}$ or $\omega_{3,0}$, were polluted by the lower order ones, even after projecting with the corresponding polynomial. As a result, the numerical results for high-order coefficients did not show the correct $r$ power law. We resolved this issue by subtracting the lower-order contributions from the appropriate metric function, and {\em then} projected with the suitable polynomial and fitted the radial part. For instance, instead of using the analytically correct expression
\begin{equation} \nu_{4,0}\sim\lim_{r\rightarrow\infty}r^{5}\int_{-1}^{1}\nu(r,\theta)P_{4}(\cos\theta)d\cos\theta, 
\end{equation}
we performed the following calculation,
\begin{equation} \nu_{4,0}\sim\lim_{r\rightarrow\infty}r^{5}\int_{-1}^1{\Tilde{\nu}}(r,\theta)P_{4}(\cos\theta)d\cos\theta, 
\end{equation}
where
\begin{equation} 
\Tilde{\nu}(r,\theta)=\nu(r,\theta)-\nu_2(r)P_2(\cos\theta)-\nu_0(r)P_0(\cos\theta).
\end{equation}
This procedure deletes the pollution in the coefficient, and allows for a proper radial power law fitting.

\subsection{Moments of inertia and differential rotation}
In contrast to stars that can be described as perfect fluids, rotating BS are very different from their static counterparts. As it is impossible to obtain slowly rotating BS as perturbations of the static solution, at least for the cases and the numerical methods used in the present paper \cite{Kobayashi:1994qi}, we require a full-rotating treatment\cite{Silveira:1995dh,Ferrell:1989kz}. BS cannot be understood as rigidly rotating systems, and in our previous work \cite{Adam:2022nlq} we showed how to obtain the moment of inertia from the Noether current. This general procedure was also used in \cite{DiGiovanni:2020ror} and does not rely on any approximation, taking advantage of the fact that there is a natural four-vector associated with the global $U(1)$ symmetry of the Lagrangian, i.e., the corresponding Noether current,
\begin{equation}
    j^{\mu}=\frac{i}{2}\sqrt{|g|}g^{\mu\nu}\left[\Phi^*\nabla_{\nu}\Phi-\Phi\nabla_{\nu}\Phi^*\right],
\end{equation}
which gives rise to the conserved particle number $N=\int  j^0 \sqrt{-g}d^3x$.
Now, we define the differential angular velocity as, 
\begin{equation}
\begin{split}
    \Omega=\frac{j^{\psi}}{j^{t}}=\frac{wg^{\psi t}-ng^{\psi\psi}}{wg^{tt}-ng^{t\psi}}=\frac{W}{r}+\frac{ne^{2(\nu-\beta)}}{r^2\left(w-\frac{nW}{r}\right)\sin^2\theta}.
\label{diferentialfrequency}
\end{split}
\end{equation}
Remarkably, the expression in \cref{diferentialfrequency} agrees with that obtained by Ryan in \cite{PhysRevD.55.6081} in the strong coupling approximation. As $\Omega$ is a function of $r$ and $\theta$, this must be taken into account when we compute the inertia tensor. Therefore, for a differentially rotating system, we use the following generalized expression,
\begin{equation}
    I=\int_0^{\pi}\int_0^{\infty}\frac{j(r,\theta)}{\Omega(r,\theta)}r^2\sin\theta e^{\nu+2\alpha+\beta}drd\theta,
\end{equation}
where $j(r,\theta)=T^t_{\psi}$ is the angular momentum density.

\section{Analysis}\label{results}

First of all, we introduce the standard definitions for the reduced multipoles \cite{2013Sci...341..365Y,Yagi:2014bxa},
\begin{equation}
\begin{split}
    &m_{2n}\equiv (-1)^n\frac{M_{2n}}{\chi^{2n}M_0^{2n+1}}\\
    &s_{2n-1}\equiv (-1)^{n+1}\frac{S_{2n-1}}{\chi^{2n-1}M_0^{2n}}.
\end{split}
\end{equation}
In our case, as we want to work with the mass $M_{99}$, which is $99\%$ of the total mass, we have to perform the change $|M_0|\rightarrow |M_{99}|$ and we recall that the dimensionless spin parameter is given by $\chi\equiv S_1/M_{99}^2$. As a result, our \emph{reduced multipole moments} are

\begin{equation}
\begin{split}
    \bar{I}=&\frac{I}{M_{99}^3},\hspace{0.2cm}\bar{Q}=m_2=\frac{M_2}{M_{99}^3\chi^{2}},\hspace{0.1cm}\chi=\frac{S_1}{M_{99}^2},\\
    &s_3=-\frac{S_3}{M_{99}^4\chi^4},\hspace{0.4cm}m_4=\frac{M_4}{M_{99}^5\chi^4}.
    \label{reduced}
\end{split}
\end{equation}

 As we did in our previous work \cite{Adam:2022nlq}, for the $\Bar{I},\chi,\Bar{Q}$, $n=1$ data, we represent our simulations in 3D spaces where the different multipole moments will play the role of the dimensions. As we will see below, we can find a given surface in each triad of moments. We also fit the surfaces finding a direct relationship between the three involved quantities. This means that having two of them, the third is determined within some error. Concretely, for the different quantities and harmonic indexes, our relations are fulfilled with an error of $\sim 7\%$ for the worst case and less than $ 2\%$ in general. For the sake of clarity, we discuss each quasi-universal behavior separately.

 \subsection{Completeness for the I-$\chi$-Q relations}
In comparison to \cite{Adam:2022nlq}, in the present paper we improved the coefficient fitting, added some models which reach secondary branches in the mass-frequency curves, and used much higher self-interaction constants for the quartic potential (all the used models are shown in  \cref{appendixA}). We have obtained the moment of inertia as a function of the spin parameter and the quadrupole moment. Still, this time we found a better surface fitting using the expression 
\begin{equation}
    \begin{split}
       &\beta=A_0+A_s^m\chi^m\left(\alpha-B\right)^s,
    \end{split} \label{fit-func} 
\end{equation}
where
\begin{equation}
  \begin{split}  
    &\beta=\sqrt[3]{\log_{10}{\bar{I}}},\\&
    \alpha=\log_{10}\bar{Q}.
    \end{split}
\end{equation}
Further, $s={1,2,3}$, $m={0,1,2}$. The fitting coefficients for all the fitting surfaces are shown in \cref{appendixC}. This strategy to take some roots of the logarithm was also used in \cite{Pappas:2013naa}.
The difference between the fitted surface and the real data is always less than $1\%$ for $n=1$. Let us compare the shapes of that surface with the one obtained in our previous work. It is clear how the additional BS models with second branches have increased the range of points in the 3D space, making the low quadrupolar momentum region quite bent. Nevertheless, all data points lie on a smooth and easy-to-fit surface, with an error below $1\%$, which allows us to reinforce our previous results.

\begin{figure*}[]
\centering
\hspace*{-0.0cm}\includegraphics[width=0.50\textwidth]{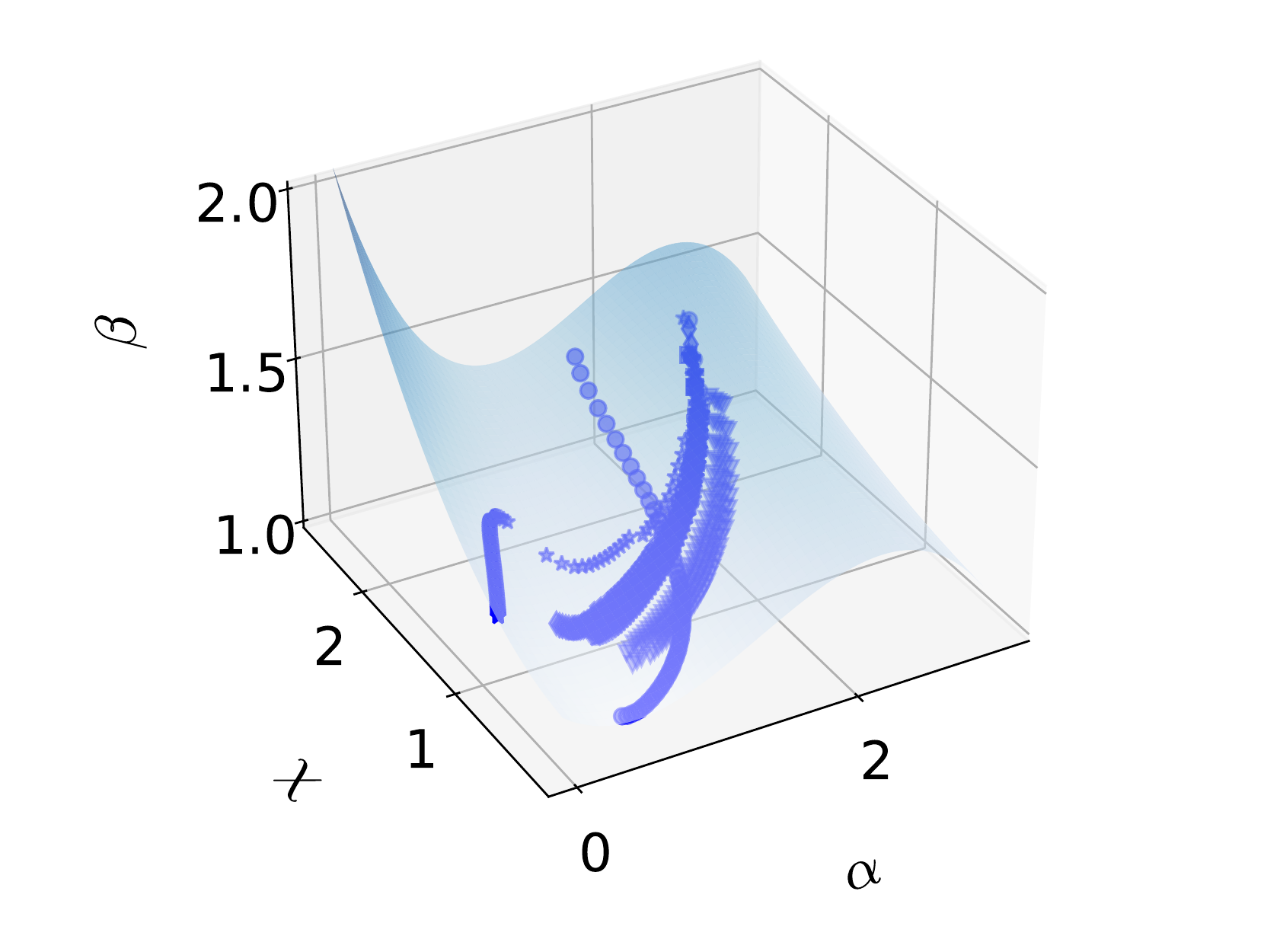}
\caption{Universal $\beta-\chi-\alpha$ surface for $n=1$ spinning BSs fitting the data points.}
\label{super}
\end{figure*}

\begin{figure*}[]
\centering
\hspace*{-0.0cm}\includegraphics[width=0.50\textwidth]{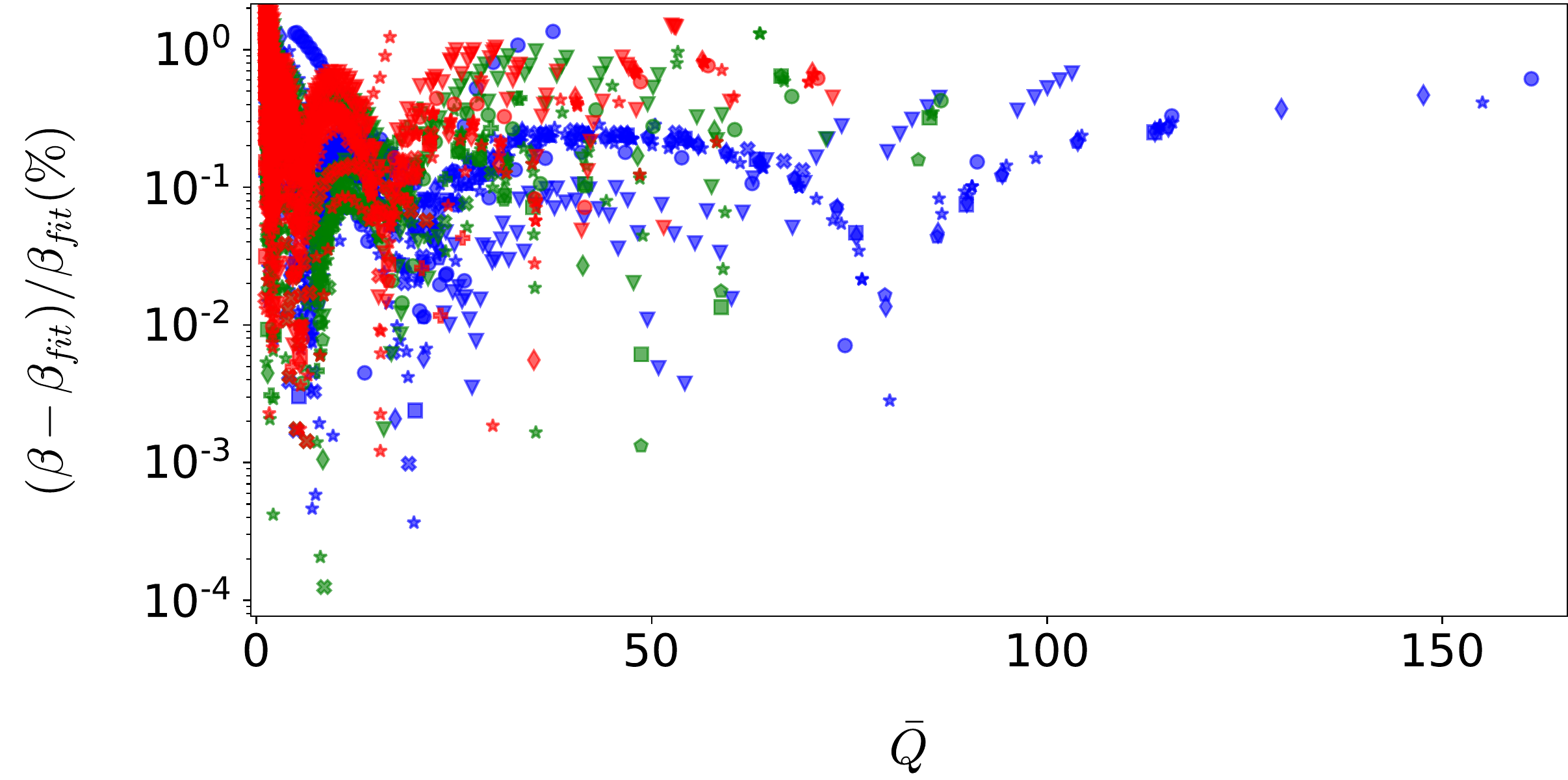}
\caption{Errors between the data and the surfaces for $n=1,2,3$ in blue, green, and red respectively. As mentioned, the maximum errors are less than $1\%$, $1\%$, and $1.5\%$ for each case. }
\label{supererror1}
\end{figure*}

For harmonic index $n=2$ and the same self-interaction models, we find that they behave similarly to the previous case.  Applying the previous techniques and the same fitting function, we find that, again, all the points lie on a smooth surface. Of course, although we use the same  fitting function \cref{fit-func}, the coefficients and the shape itself are different. As we can read from \cref{supererror1}, the highest deviation between the simulated data and the  surface is $\sim 1\%$ at the maximum. It is also clear that the $n=1$ and $n=2$ stars lie on two distinct surfaces.

\begin{figure*}[]
\centering
\hspace*{-0.0cm}\includegraphics[width=0.50\textwidth]{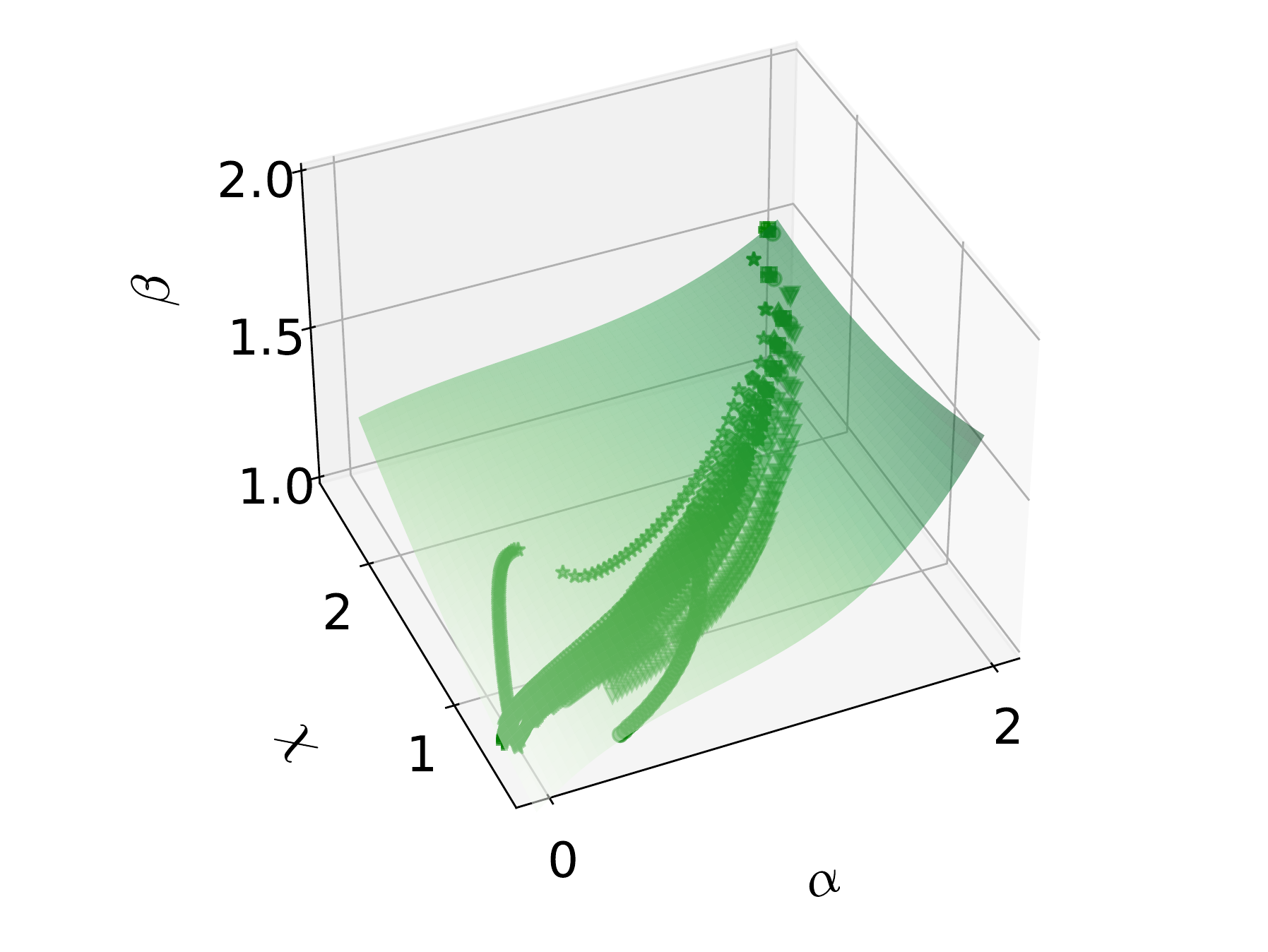}
\caption{Universal $\beta-\chi-\alpha$ surface for $n=2$ spinning BSs. The relation holds with an error of less than $1\%$.}
\label{super2}
\end{figure*}

\begin{figure*}[]
\centering
\hspace*{-0.0cm}\includegraphics[width=0.50\textwidth]{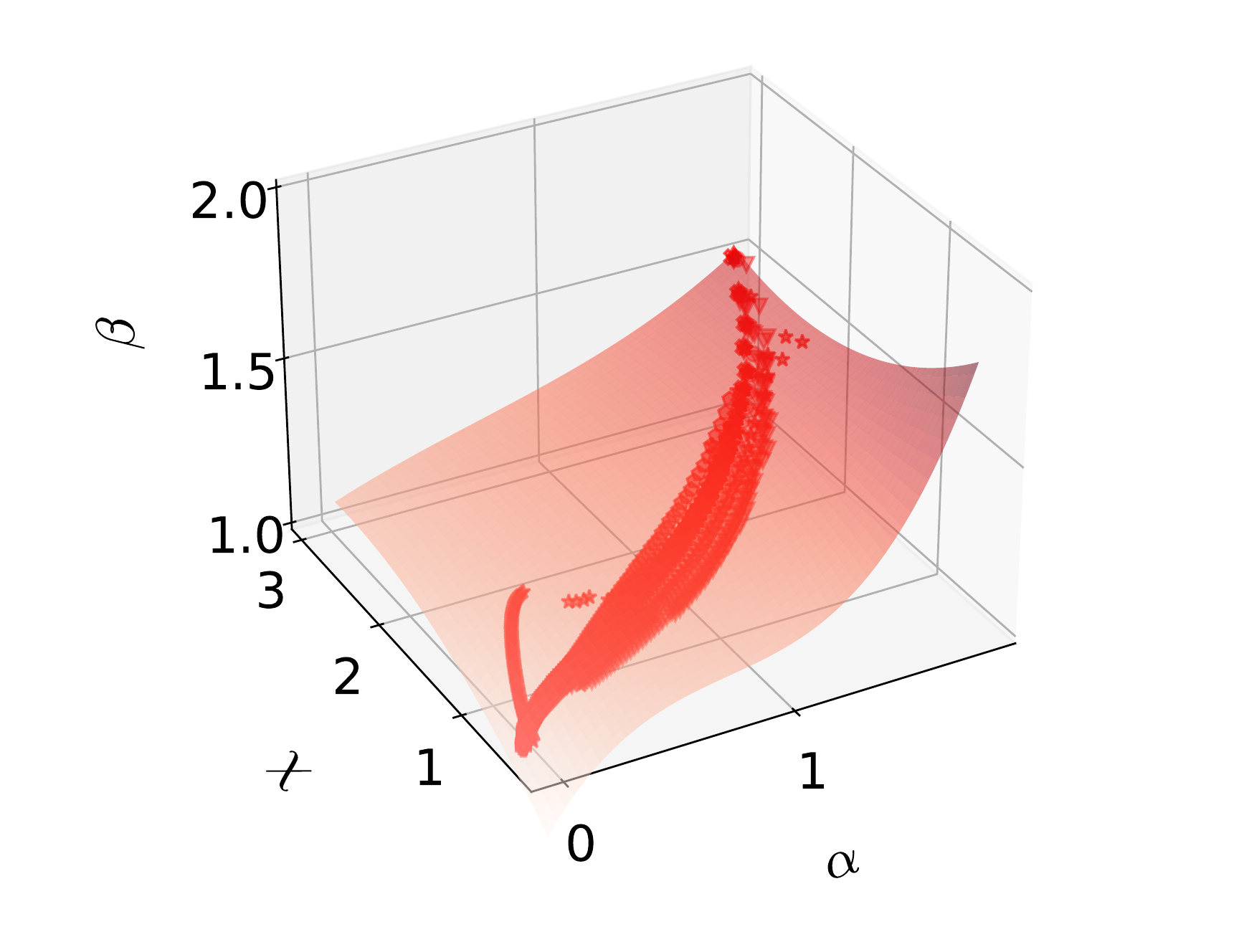}
\caption{$\beta-\chi-\alpha$ surface for  $n=3$ spinning BSs. The relation holds with an error of less than $1.5\%$.}
\label{super3}
\end{figure*}

We also did the simulation for the harmonic index $n=3$ and applied the same analysis. We read from \cref{supererror1} that the errors in the fitting are always below $1.5\%$, so we can ensure that our stars fulfill a universal behavior, and our results given in \cite{Adam:2022nlq} can be expanded for higher harmonic indexes.

We also have observed that for higher the winding number $n$ the distance between neighboring surfaces gets closer. So $n=2$ and $n=3$ stars could be treated in a unique surface, and the distance between the  surface and points would correspond to an error of less than $5\%$, which is still a good precision in astrophysical terms. Doing so, some points can belong to any of the surfaces, because in the small region where the  moment of inertia, the quadrupolar, and the angular moment are low, the three surfaces tend to join. This is more noticeable for $n=2$  \cref{super2} and $n=3$ \cref{super3}, for which the degenerated region is larger. It is also predictable that for $n>3$, the subsequent surfaces will be closer and closer. But we have two arguments for justifying the separated fits. The first is that by doing so the errors are much lower. The second concerns the range in the dimensionless spin each model can reach. We can read from the data that the bigger the winding number, the bigger the spins. As we cannot reach a BS with $\chi>2.6$ with $n=1$, but it is relatively easy for $n=2$ stars, splitting the fitting functions by the winding numbers makes sense.

As a conclusion to this section, we have found a universal relation for each winding number in such a way that the quadrupolar moment and the angular and mass moments determine the moment of inertia with very high precision in a model-independent fashion.

\subsection{Spin octupole relations}
The higher-order multipoles for rotating BS were studied in the seminal paper by Ryan \cite{Ryan:1996nk} and recently by \cite{Vaglio:2022flq} for the quartic potential within the strong coupling constant approximation. We are not using data in that regime for the quartic potential, because we will study the high-coupling constant regime  in a future publication. Also, some comparisons are not straightforward due to our different numerical approaches. But still, we did compare our current data with \cite{Vaglio:2022flq}, with positive conclusions. Even for different models and coupling constants, the comparable regions behave similarly in tendency and values for the spin octupole $s_3$ against $\chi$.
We have found that a rescaling of this multipole with the dimensionless quadrupole moment leads to a better fit. We will, therefore, use the following redefinition of the spin octupole,
\begin{equation}
    \Bar{s}_3=\frac{s_3}{\Bar{Q}}
\end{equation}

Following previous work about the universal behavior for spinning NS \cite{Yagi:2014bxa}, we studied the possible existence of some $2D$ relations for $\Bar{s}_3$ in our BS framework. We have found that this is not the case. For some well-posed potentials included in the analysis, it was impossible to adjust any $2D$ curve with a reasonable precision, so we treat $\Bar{s}_3$ in the same fashion as before. We plot our stars in a $\Bar{s}_3-\chi-\Bar{Q}$ space, and again they form smooth surfaces.

\begin{figure*}[]
\centering
\hspace*{-0.0cm}\includegraphics[width=0.50\textwidth]{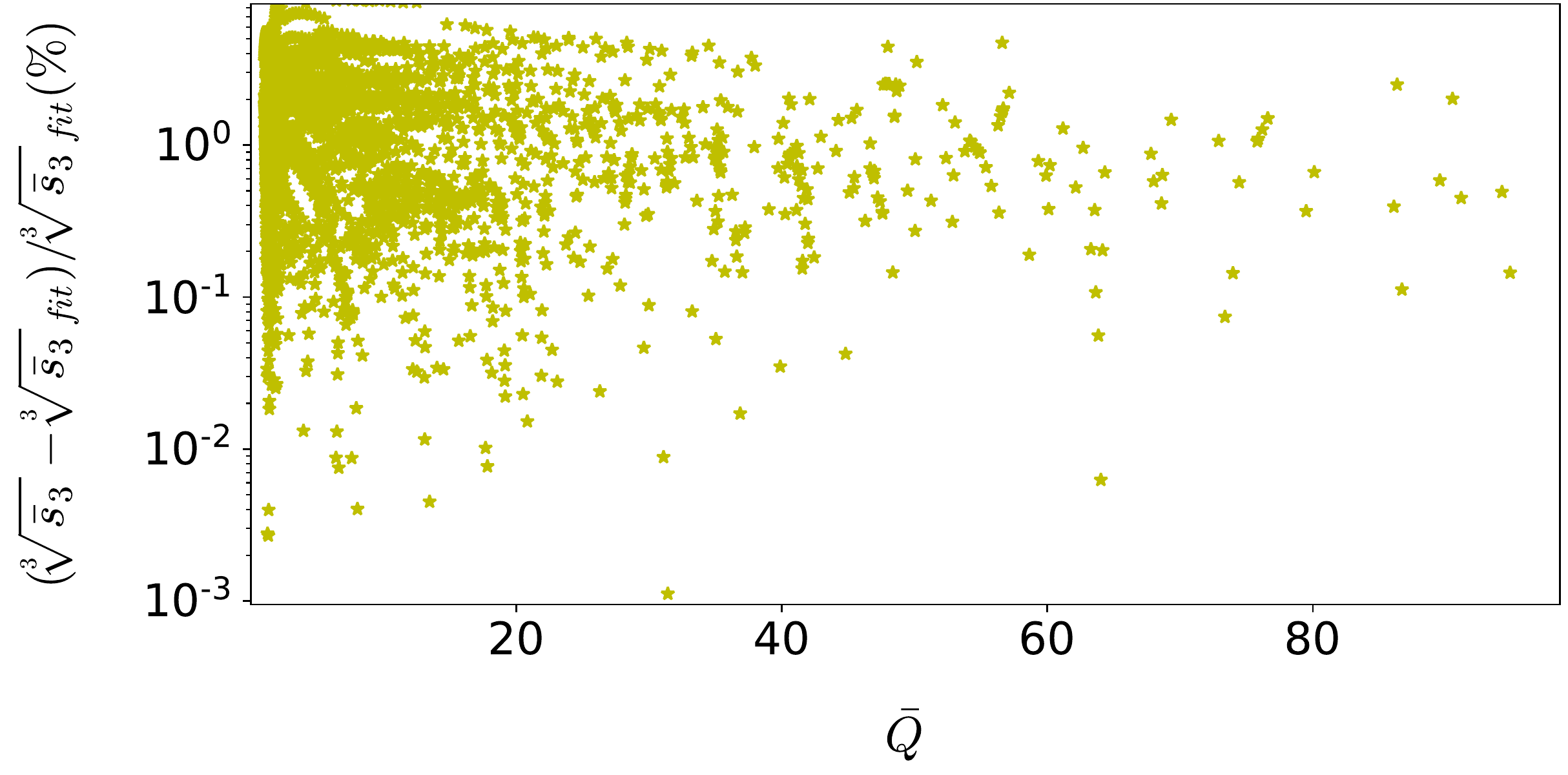}
\caption{Errors between the data and the $\sqrt[3]{\Bar{s}_3}$ surface fitting for $n=1,2,3$ stars together.  }
\label{s3xq_tot}
\end{figure*}

First, in \cref{s3xq_tot} all $n=1,2,3$ cases are plotted in one error plot. We see that all the stars can be fitted by a unique surface but in a non-satisfactory  manner, with errors of the order $10\%$. The fittings can  be significantly improved by splitting in winding number. Doing so, for each $n$, we have a different surface.
That time we use the following function,
\begin{equation}
   \sqrt[3]{\Bar{s}_3}=A_0+A_s^m\chi^m\left(\Bar{Q}-B\right)^s10^{-2s+1},
\end{equation}
where we now have $s=1,2,3,4$ and $m=0,1,2$. Further, here and in the following we extract some $s$-dependent powers of $10$, such that the resulting fitting constants $A_s^m$ are of the same order of magnitude.

\begin{figure*}[]
\subfloat{%
  \hspace*{1.0cm}\includegraphics[clip,width=1.0\columnwidth]{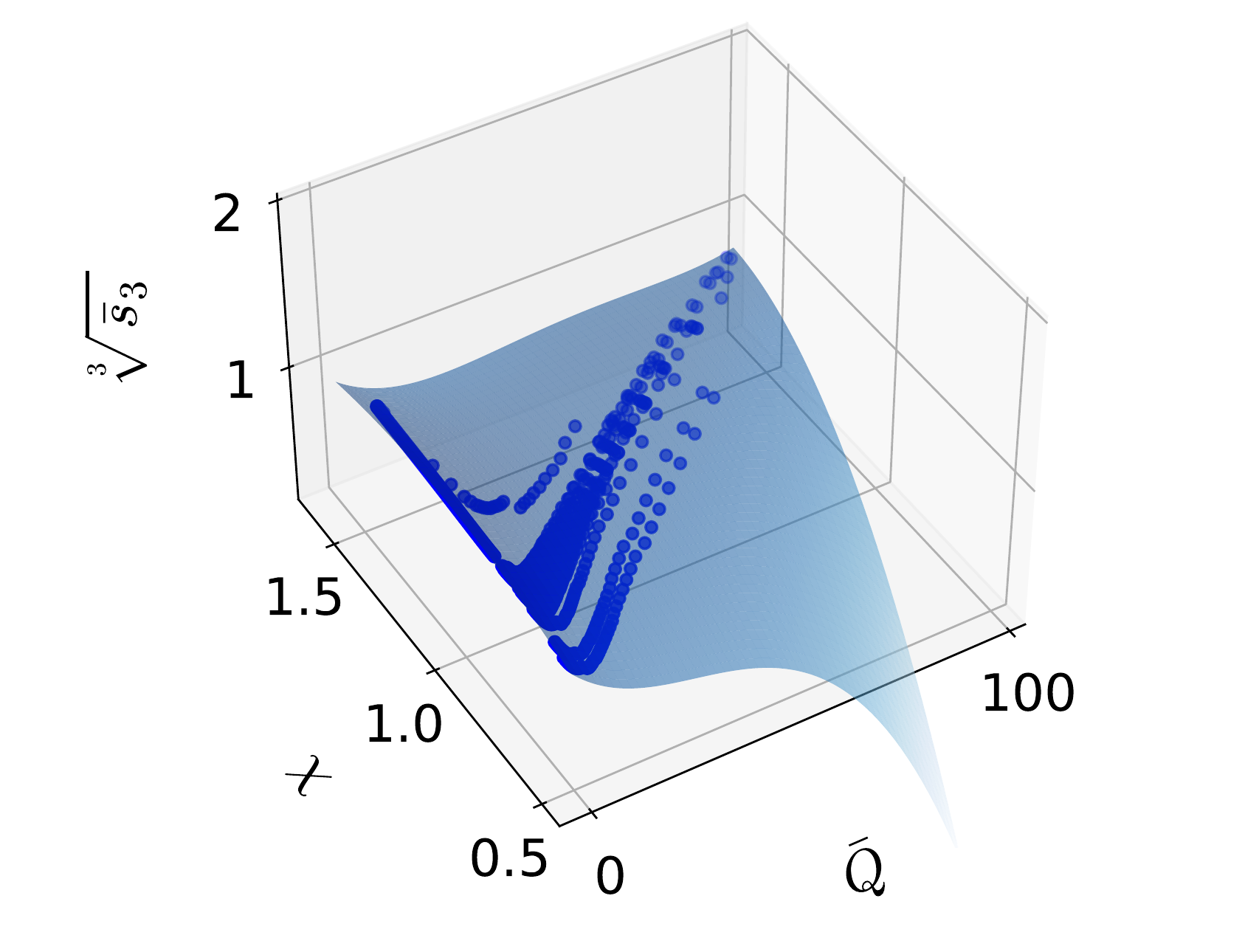}%
}

\subfloat{%
  \hspace*{1.0cm}\includegraphics[clip,width=0.94\columnwidth]{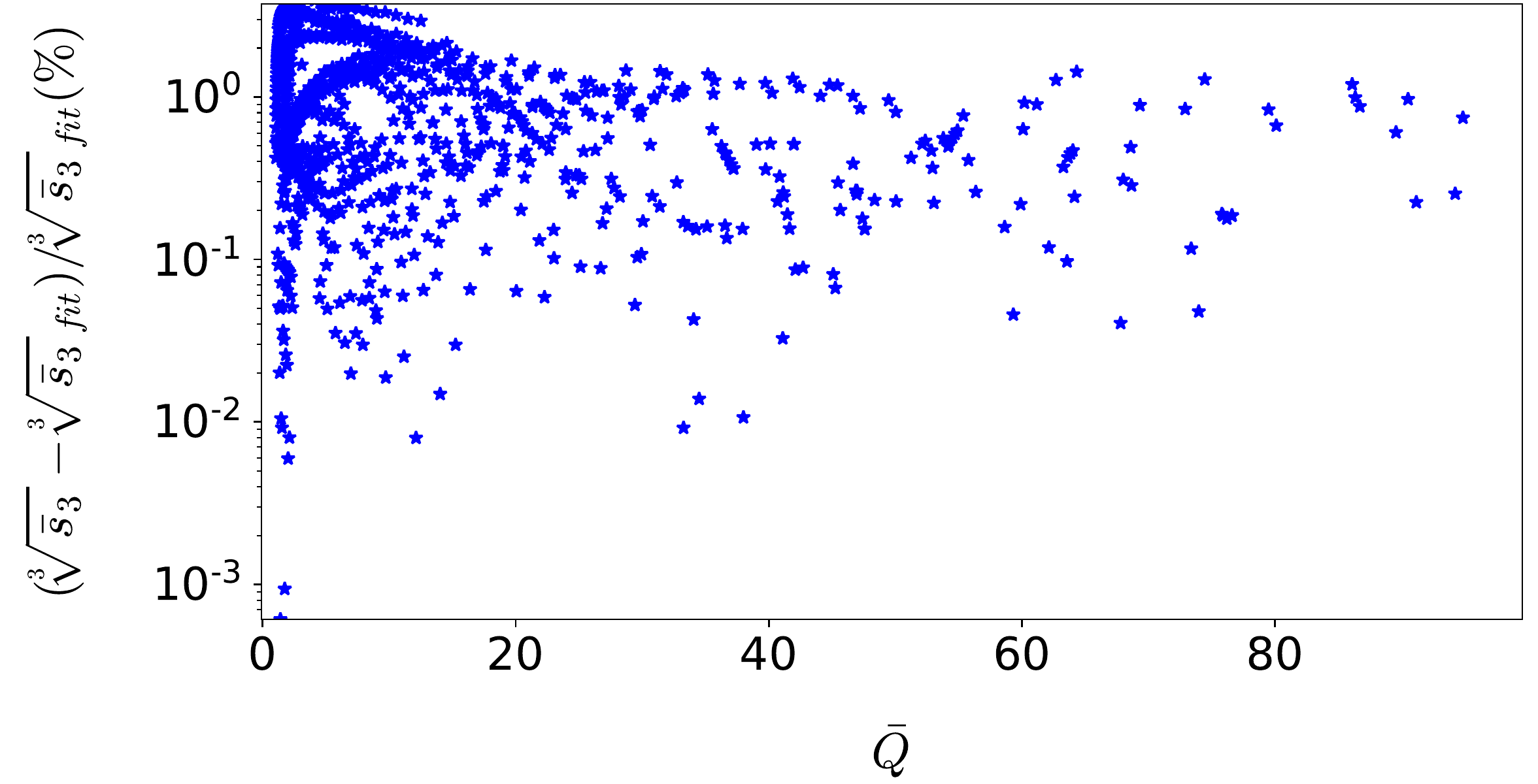}%
}
\caption{$\sqrt[3]{\Bar{s}_3}-\chi-\Bar{Q}$ surface for $n=1$ spinning BSs fitting the data points (upper panel) and the relative difference between data and fitted value (lower panel). }
\label{s3n1}
\end{figure*}

Starting with $n=1$, it can be clearly seen in  \cref{s3n1} that the stars define a smooth surface with a fitting error of less than $3.5\%$, which is sufficiently precise to assure the existence of a new kind of universal behavior in terms of the above quantities. We have also checked that in the region of very compact stars, where the quadrupolar moment is small, we find a limiting value for $s_3 = \bar{Q}\, \Bar{s}_3$  equal to the Kerr-Black-Hole limit $s_3^{KBH}=1$, as expected. 

\begin{figure*}[]
\subfloat{%
  \hspace*{1.0cm}\includegraphics[clip,width=1.0\columnwidth]{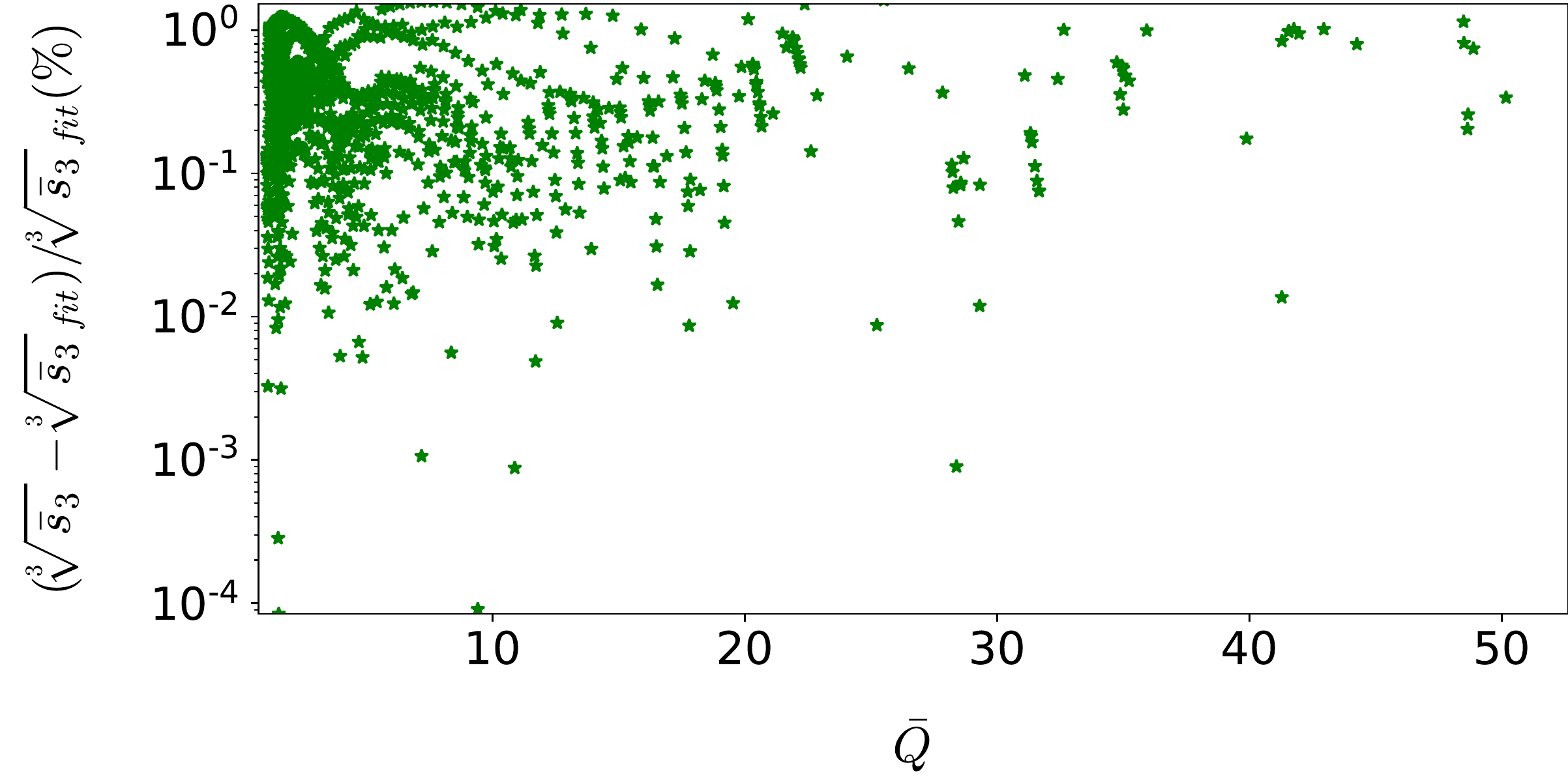}%
}

\subfloat{%
  \hspace*{1.0cm}\includegraphics[clip,width=1.0\columnwidth]{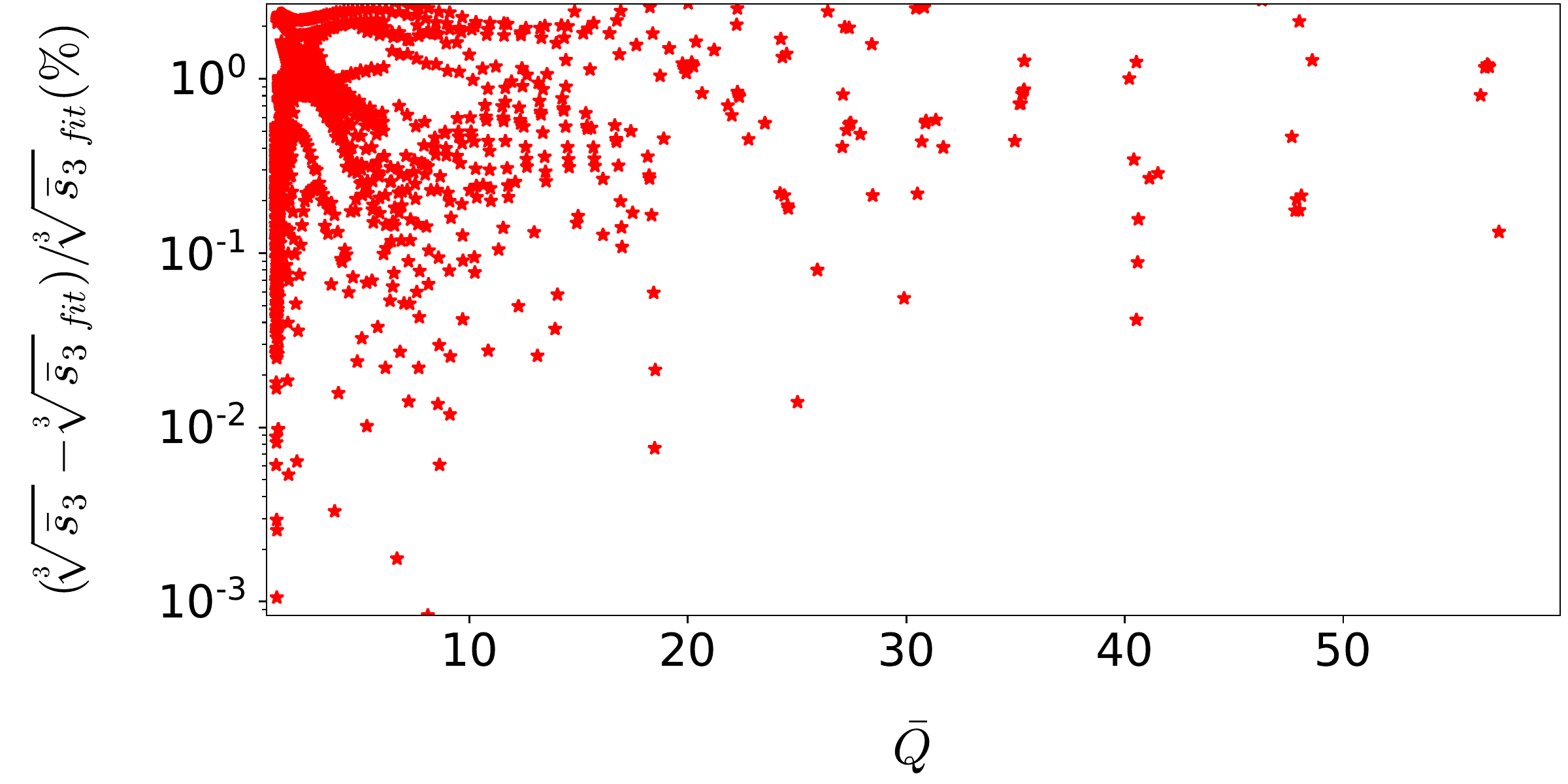}%
}
\caption{Differences between data and fittings for $n=2$ (upper panel) and $n=3$ (lower panel). }
\label{s3n23}
\end{figure*}

The same analysis for $n=2$ leads to a fitting function that is even better, allowing for a maximum error of less than $1\%$  and maintaining the universality for this harmonic index.
For $n=3$, the errors are lower than $2\%$.
So we could extract $s_3$ just by knowing $\chi, Q$ and guessing $n$ in terms of $\chi$. Both error analyses are presented in \cref{s3n23}.

\subsection{Mass hexadecapole relations}
\begin{figure*}[]
\subfloat{%
  \includegraphics[clip,width=1.0\columnwidth]{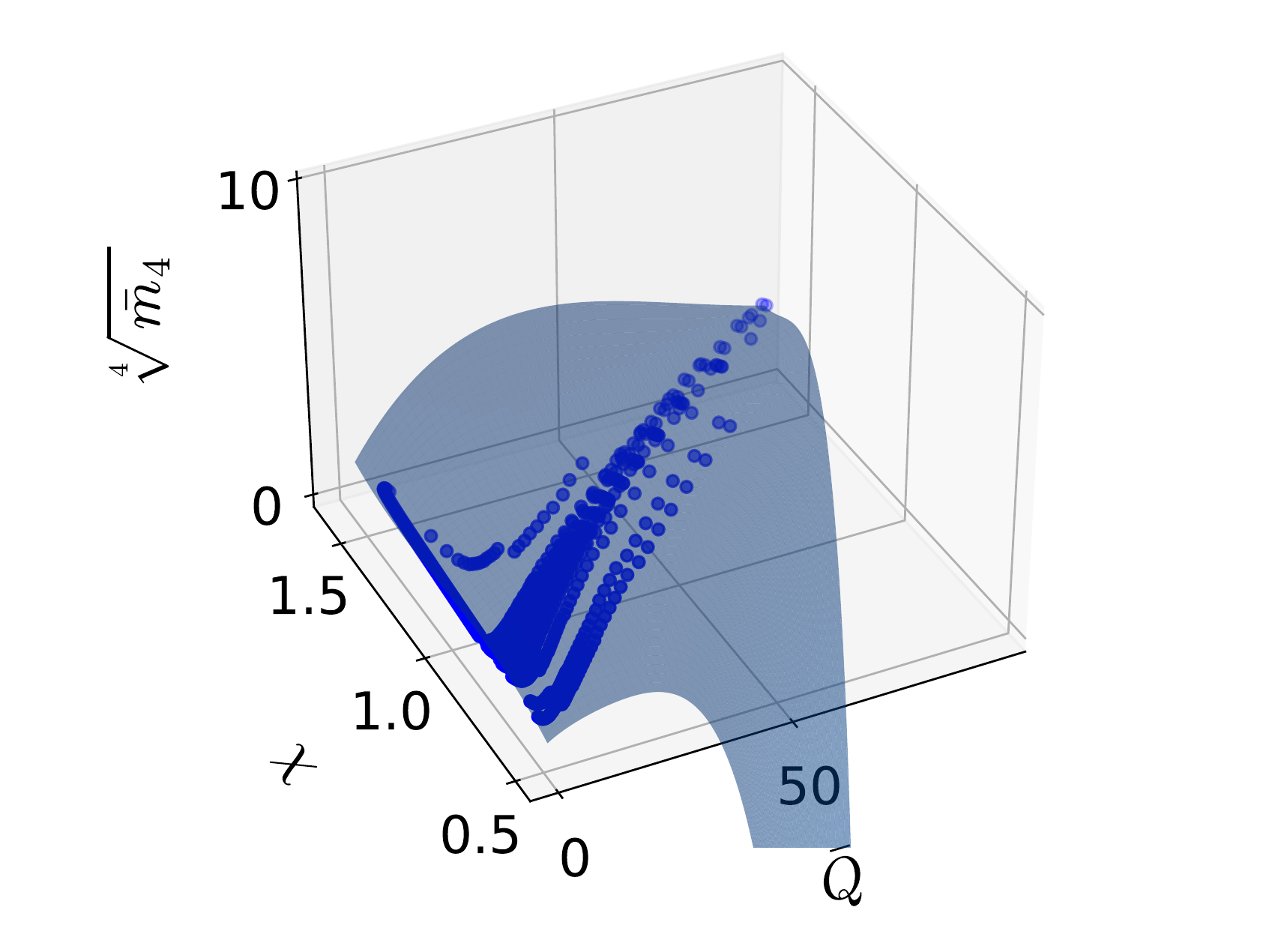}%
}

\subfloat{%
  \includegraphics[clip,width=0.94\columnwidth]{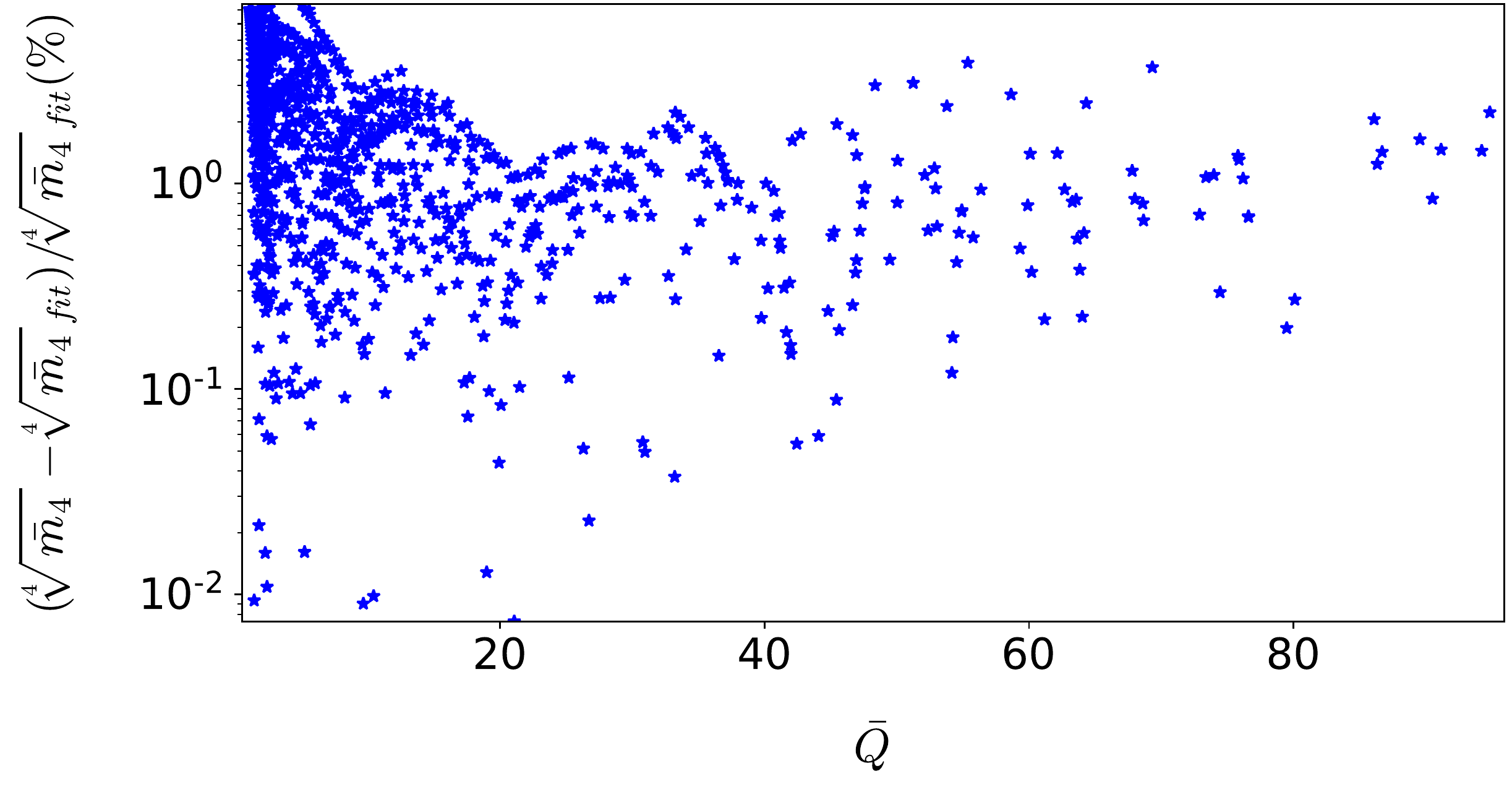}%
}
\caption{$\sqrt[4]{m_4}-\chi-\Bar{Q}$ surface for $n=1$ spinning BSs fitting the data points (upper panel) and the relative difference between data and fitted value (lower panel). }
\label{m4n1}
\end{figure*}

\begin{figure*}[]
\subfloat{%
  \includegraphics[clip,width=1.0\columnwidth]{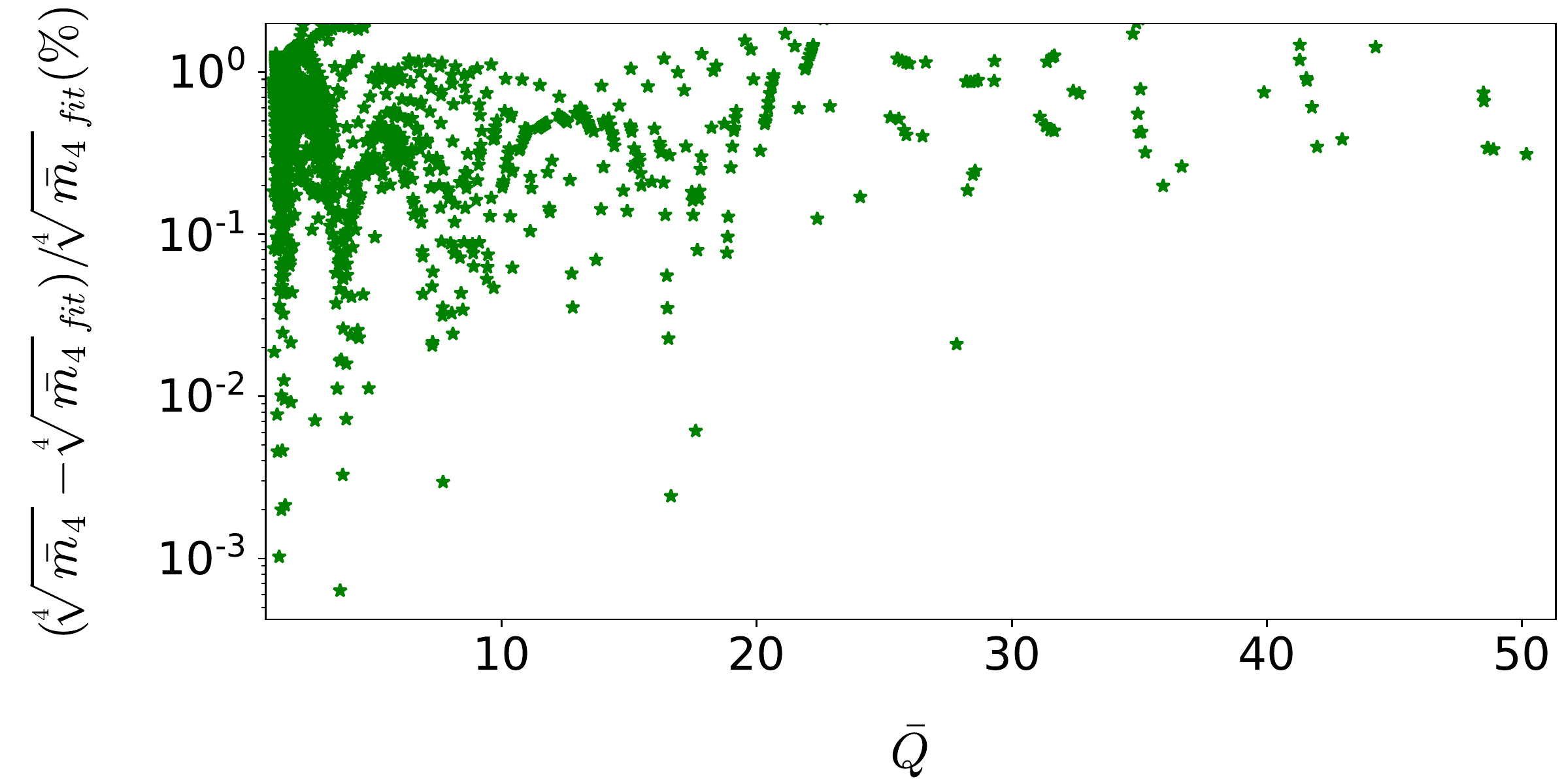}%
}

\subfloat{%
  \includegraphics[clip,width=1.0\columnwidth]{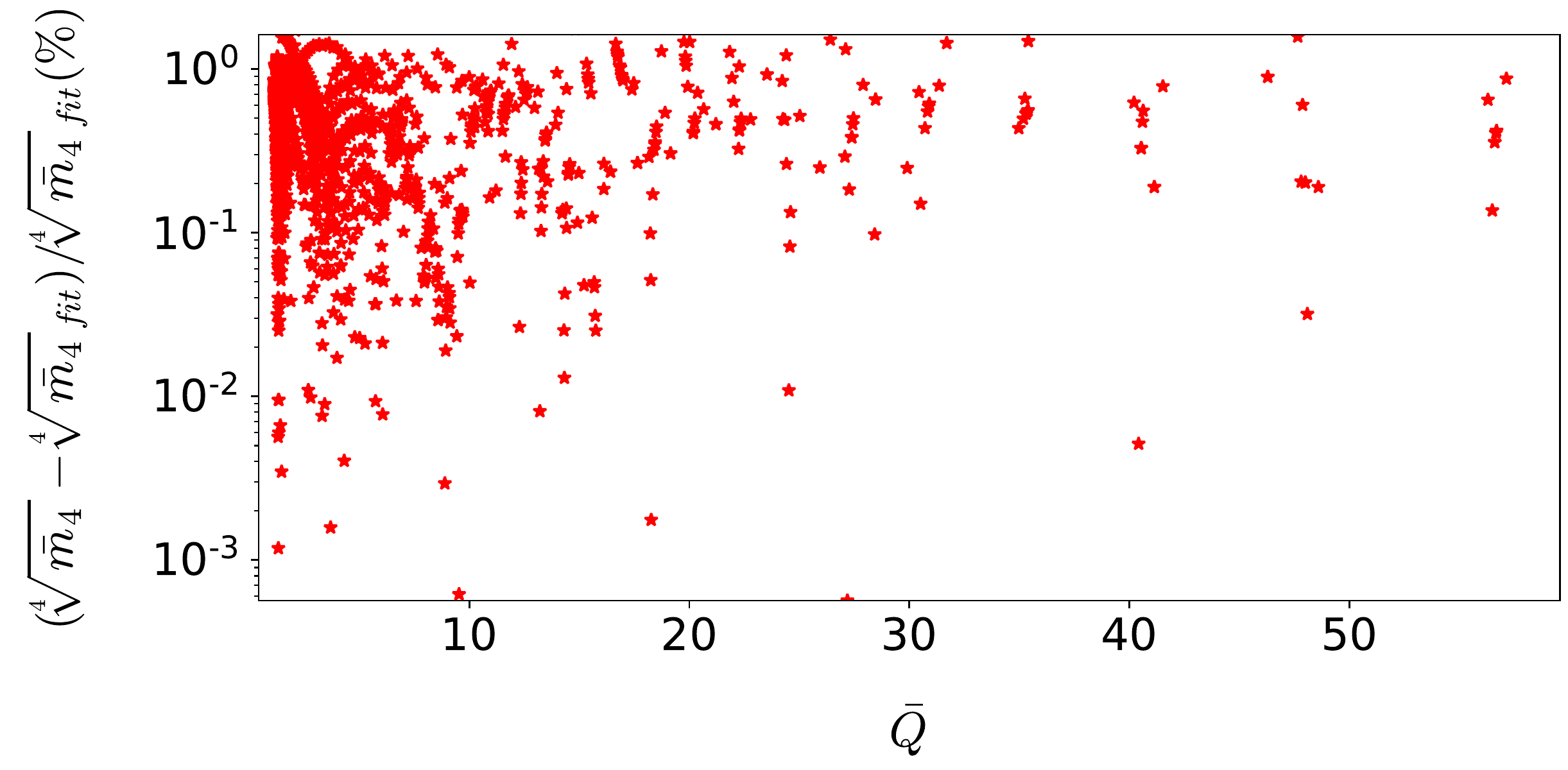}%
}
\caption{Differences between data and fittings for $n=2$ (upper panel) and $n=3$ (lower panel). }
\label{m4n23}
\end{figure*}

For the highest multipole we study in this work, the mass hexadecapole $m_4$, again, we could not find $2D$ relations. But as in the previous cases, we could find a new effective no-hair relation between the three magnitudes $m_4-\chi-\Bar{Q}$. Our fitting function for them is
\begin{equation}
   \sqrt[4]{m_4}=A_0+A_s^m\chi^m\left(\Bar{Q}-B\right)^s10^{-2s+2},
  \label{fitm41} 
\end{equation}
where $s=1,2,3,4$ and $m=0,1,2,3$. This means that more coefficients are needed to have a good fitting. 
For $n=1$, see \cref{m4n1}, our precision is lower than the previous cases; we have found that some stars have an error $\sim 7\%$  in the worst scenario. 
Surprisingly, for $n=2$ and $n=3$, the highest errors are smaller, namely $1.4\%$ and $1\%$, respectively, see \cref{m4n23}.

\subsection{Compactness}
Historically, one of the first quantities participating in universal relations  was compactness. Compactness is defined as
\begin{equation}
    \textit{C}=\frac{M}{R},
\end{equation}
where $M$ and $R$  are the mass and radius of a star, and their precise definition may slightly vary, depending on the astrophysical object under consideration.
The first universal behavior for compact stars was found during the 90s
\cite{Lattimer:1989zz}, linking the binding energy of NS with their compactness.

Also, some studies found relations between the $f-mode$ and $w-mode$ frequencies, and the compactness $\textit{C}$ \cite{Andersson:1996pn,Andersson:1997rn}. Some relations were even found between the damping time for both modes and, again, the compactness.

More recently, in the 2000s, relations between $\textit{C}$ and the NS moments of inertia have been found and studied \cite{Lattimer:2000nx,Lattimer:2004nj,Bejger:2002ty}
Further, relations between the compactness and the quadrupole moments were found in \cite{Urbanec:2013fs}.

The discovery that the compactness was directly related both to the moment of inertia and the quadrupolar moment, led to the study of the relations between these two quantities. Then the field increased further, leading to a large number of neutron and quark star universal relations \cite{Doneva:2017jop,Yagi:2016bkt,Yagi:2014bxa,Yagi:2013awa}.

\begin{figure*}[]
\subfloat{%
  \hspace*{1.0cm}\includegraphics[clip,width=1.0\columnwidth]{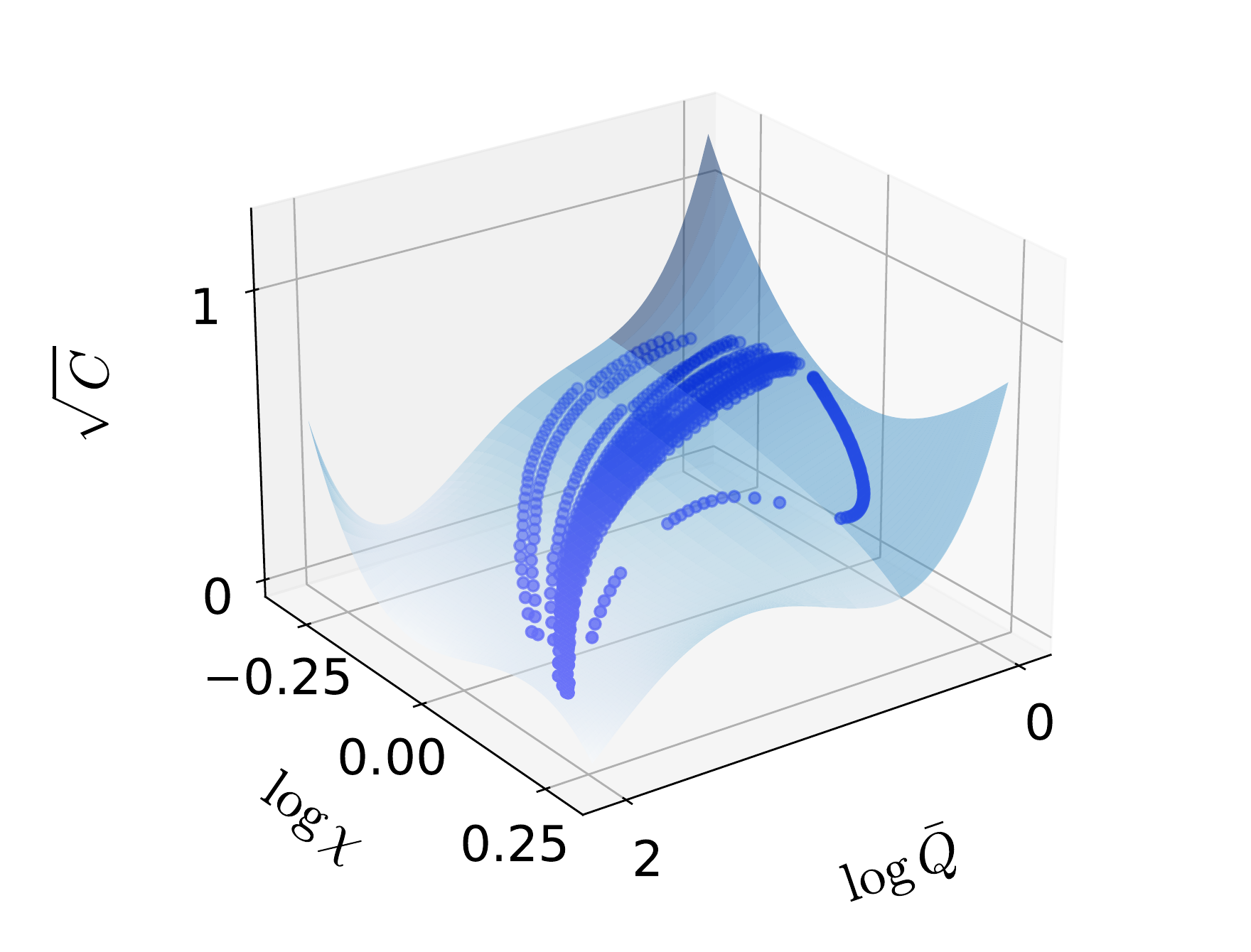}%
}

\subfloat{%
  \hspace*{1.0cm}\includegraphics[clip,width=1.0\columnwidth]{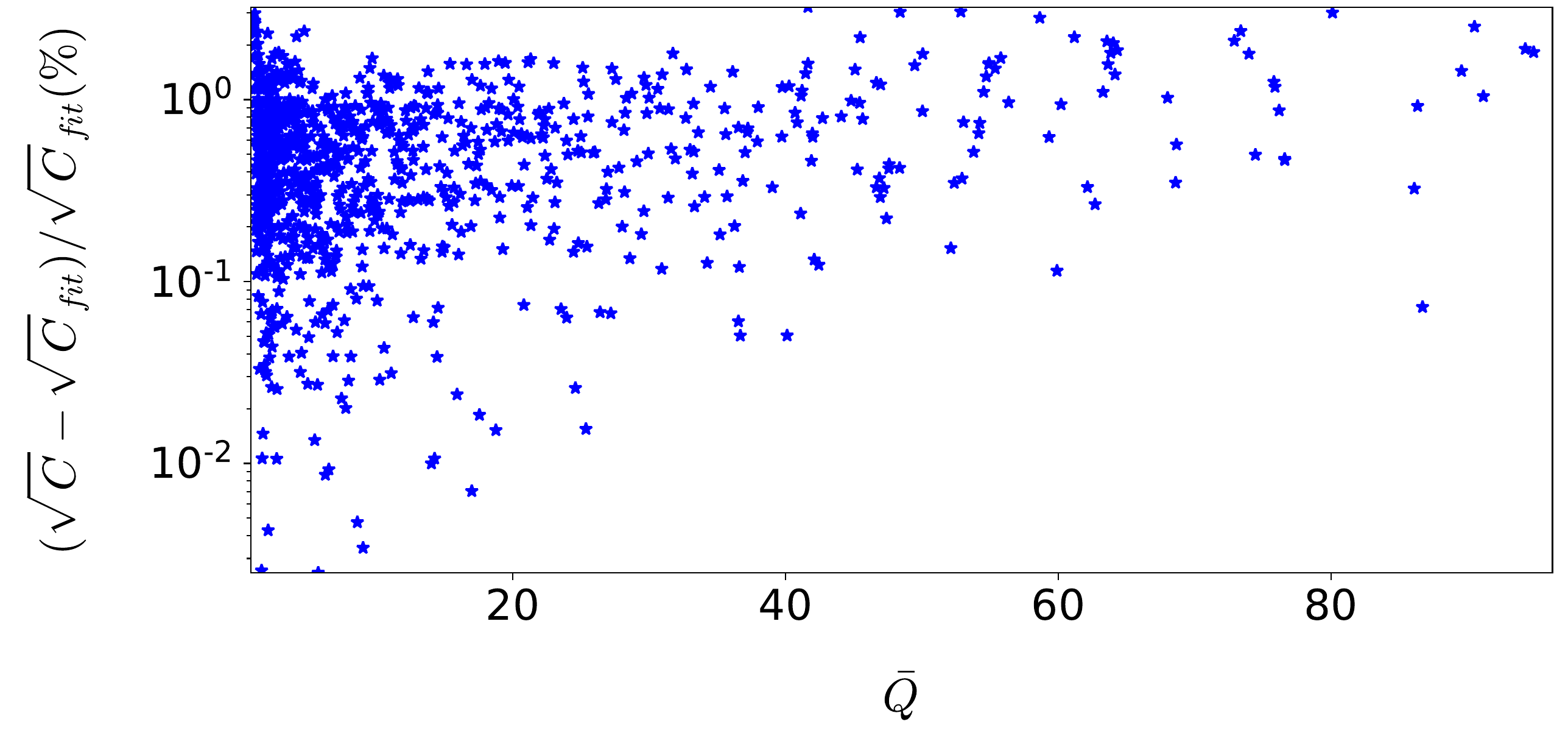}%
}
\caption{$\sqrt[2]{\textit{C}}-\log_{10}\chi-\log_{10}\Bar{Q}$ surface for $n=1$ spinning BSs fitting the data points (upper panel) and the relative difference between data and fitted value (lower panel). }
\label{cn1}
\end{figure*}

\begin{figure*}[]
\subfloat{%
  \hspace*{1.0cm}\includegraphics[clip,width=1.0\columnwidth]{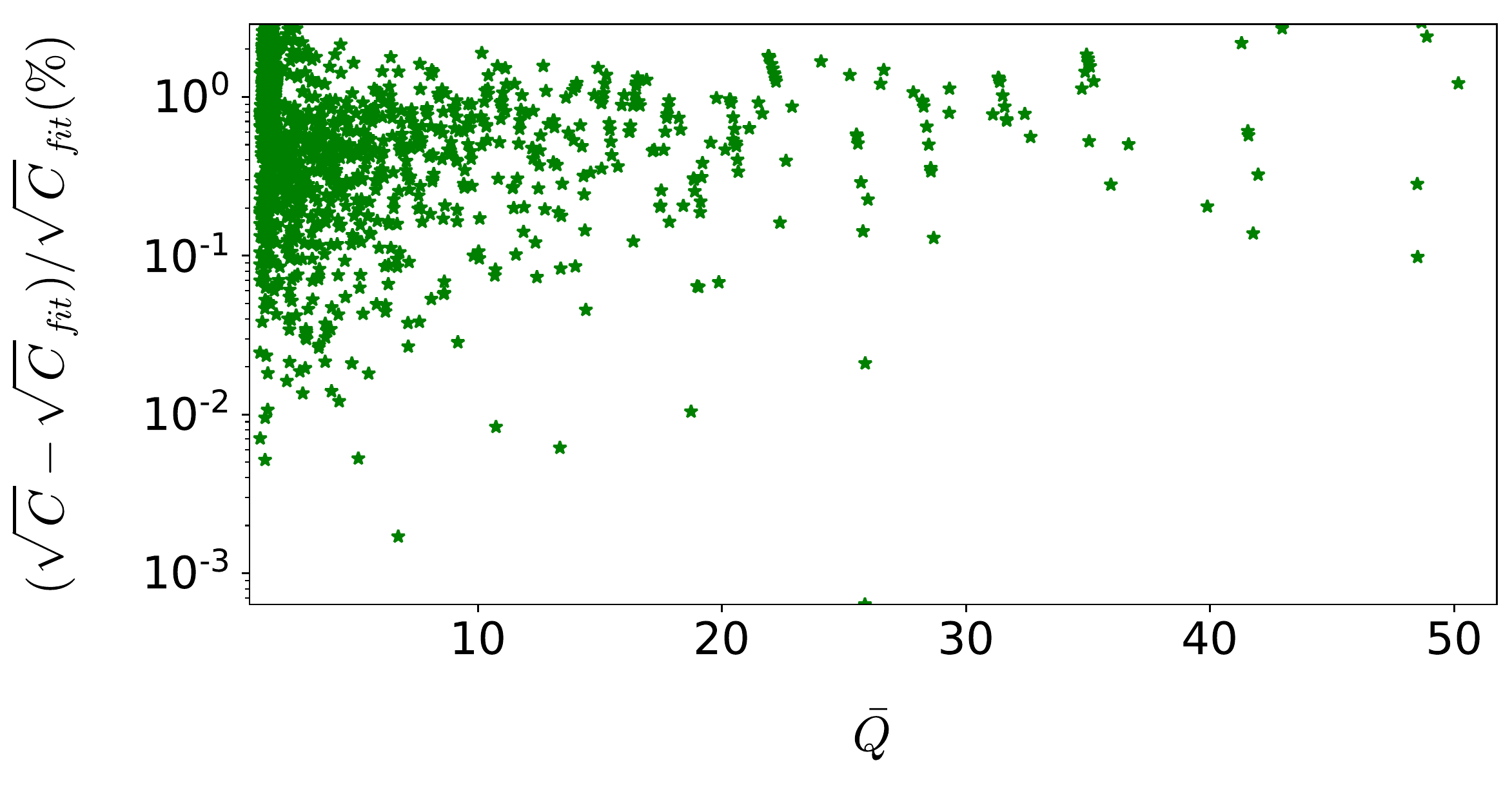}%
}

\subfloat{%
  \hspace*{1.0cm}\includegraphics[clip,width=1.0\columnwidth]{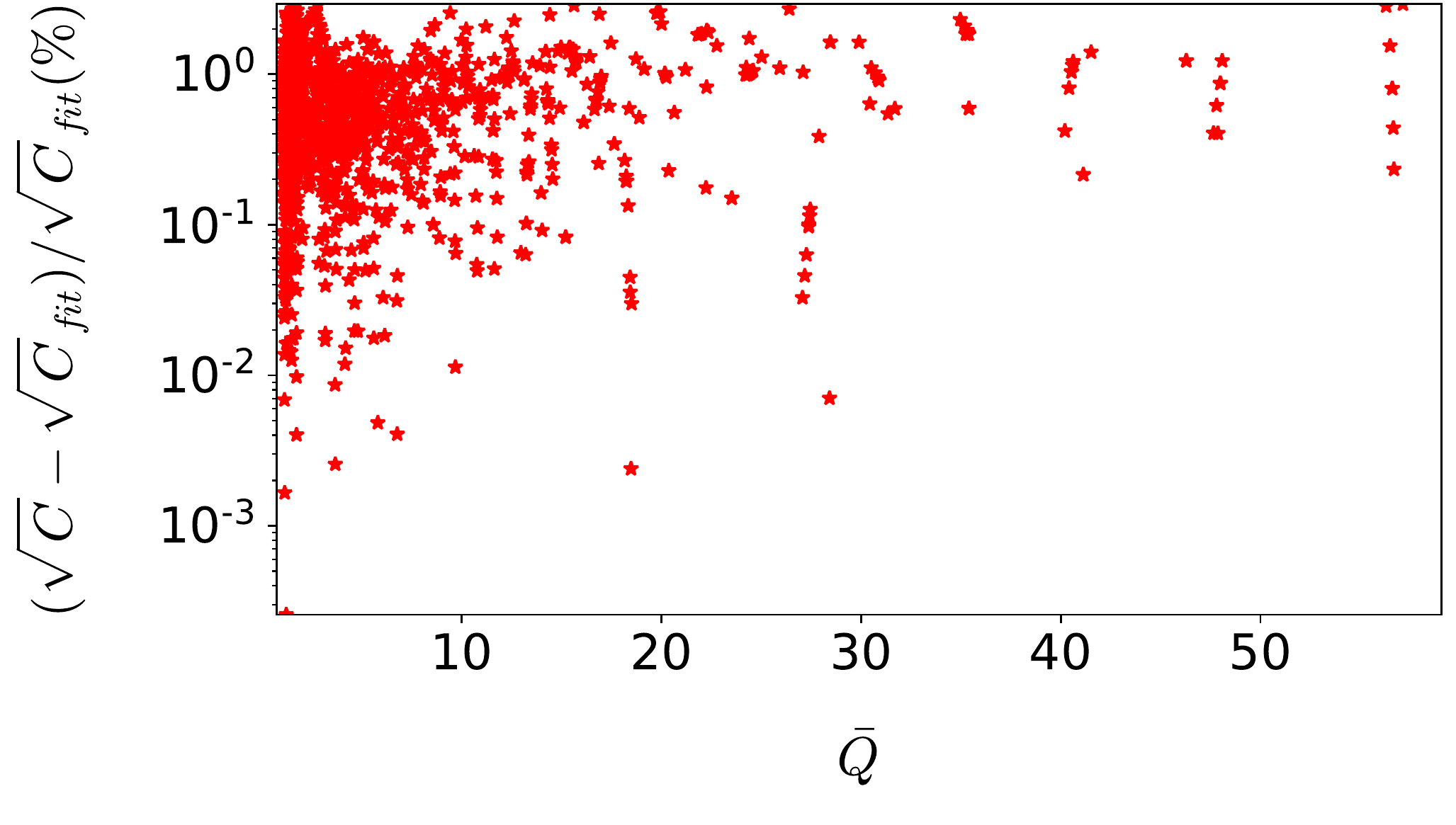}%
}
\caption{Differences between data and fittings for $n=2$ (upper plot) and $n=3$ (lower one). }
\label{cn23}
\end{figure*}

Within the BS paradigm, the universal relations and effective no-hair properties were not studied as widely, because these stars are more exotic objects. But apart from our previous work, the topic was discussed in \cite{Ryan:1996nk,Vaglio:2022flq,Grandclement:2014msa}. 
Focusing on the compactness, we can compare our data with \cite{Vaglio:2022flq}, and we see that despite the differences between models and regimes, our results seem to agree. We expected and eventually found a new relation linking compactness and different multipoles. In \cref{cn1}, we show our BS in a $\textit{C}-\log\Bar{\chi}-\log\bar{Q}$ parameter space, and it is clearly visible that all data form a smooth surface with a less than $2.5\%$ error deviation, for $n=1$ stars, \cref{cn1}. Again, we can fit our data with a lower than $3\%$ deviation for higher harmonic indexes, $n=2$ and $n=3$, see \cref{cn23}. This clearly demonstrates the universal behavior for compactness. Like in the previous cases, we could fit all $n=1,2,3$ stars together, but the results would lose accuracy.
The fitting procedure was done using the function
\begin{equation}
   \sqrt[2]{\textit{C}}=A_0+A_s^m\log_{10}\chi^m\left(\log_{10}\Bar{Q}-B\right)^s,
  \label{fitm4} 
\end{equation}
with $s=1,2,3,4$ and $m=0,1,2,3$.

An interesting observation is that contrary to the NS case, we could not fit to a proper smooth surface with any other triplet of parameters, consisting of the compactness and two more ($m_4,s_3,I...$, or even the field frequency $w$). More concretely, some low-frequency stars which we could include in the previous sections, make impossible other quasi-universal relations, even for the most common potentials or regimes. As we want to be as general as possible, we, therefore, cannot show any other universal or quasi-universal behavior. It is also interesting that, although we use a slightly different compactness definition in comparison to \cite{Vaglio:2022flq} - since we are using the radius $R_{99}$ that encloses the mass $M_{99}$ - we found approximately the same limit for the maximum compactness value, that is, $\textit{C}_{max}\sim 0.4$ in our current data set.

\subsection{Comparison and comments concerning NSs and BSs}

It is worth pointing out that in the NS framework, we have enough freedom to fix the mass and $\chi$ independently, while for BS, one of the two fixes the other. This means that the moments of inertia of NS solutions span a   $(\chi, \bar{Q})$ surface for a given model, while for BS, they follow curves. So universality comes from the fact that all data lie on the same surface, independent both of the model and the coupling constants.

\begin{figure*}[]
\centering
\hspace*{-0.0cm}\includegraphics[width=0.60\textwidth]{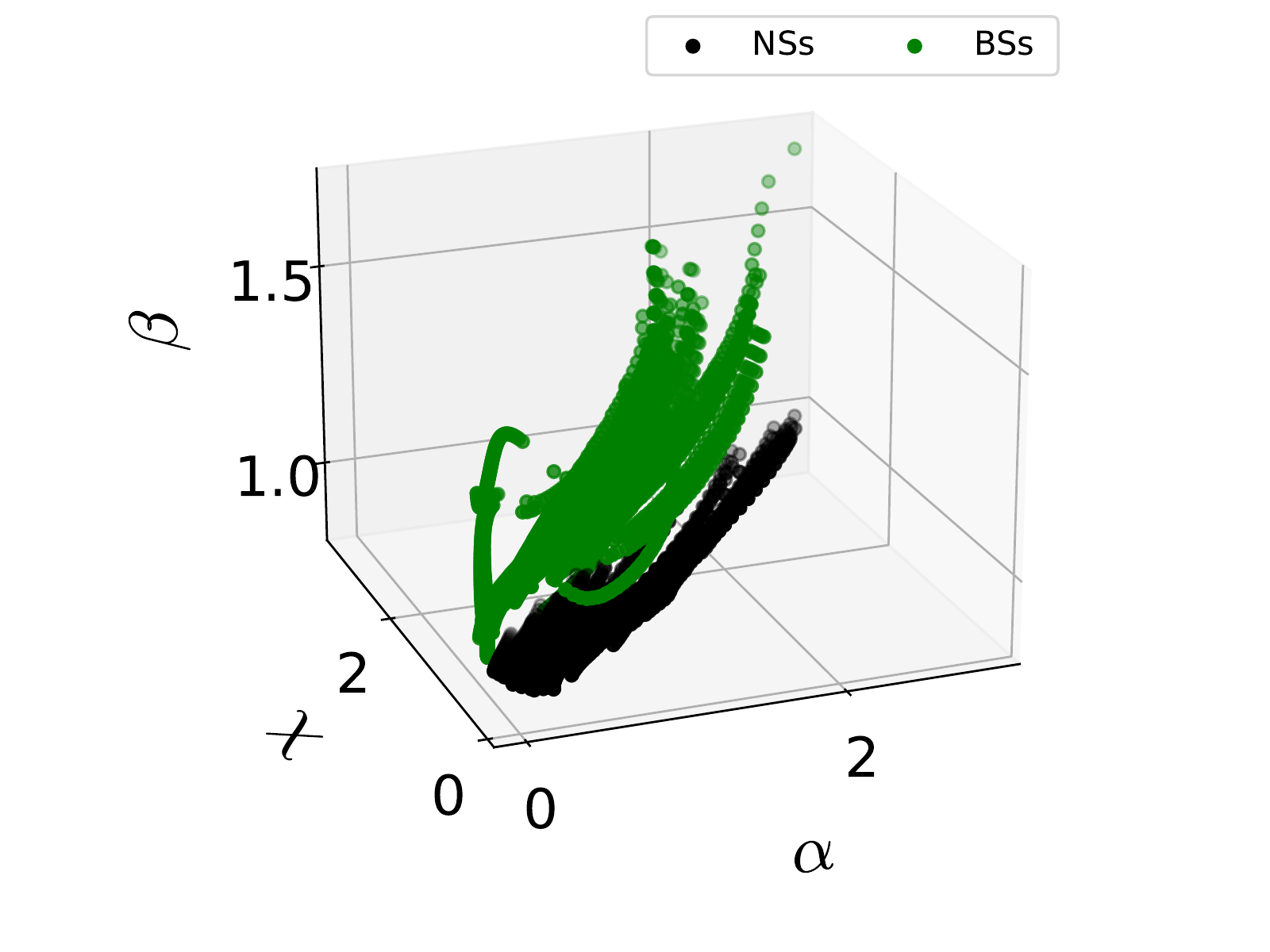}
\caption{Green dots correspond to $n=1,2,3$ BS data. Black points are NS for different frequencies and several EOS, namely BCPM \cite{Sharma:2015bna}, AGHV \cite{Adam:2020yfv}, BPAL \cite{Zuo:1999vcl}, RNS-FPS \cite{Engvik:1994tj}, RNS-A \cite{Arnett:1977czg} and SLy \cite{Douchin:2001sv}. }
\label{NS}
\end{figure*}

An interesting comparison can be made between our fitted surface for rotating BS and a similar result for rapidly rotating NS. As shown in \cite{Pappas:2013naa,Yagi:2014bxa}, the NS moments of inertia, spin parameter, and quadrupole moments  fulfill some universal relations. We plot NS and BS data together in the same parameter space $\beta-\chi-\alpha$, being $\beta=\sqrt[3]{\log_{10}\Bar{I}}$ and $\alpha=\log_{10}\Bar{Q}$.  Rapidly rotating NS data were obtained using the RNS package  \cite{stergioulas1992rotating}. In \cref{NS}, black dots are rotating NS for various EOS, ranging from low to mass shedding velocities, while green dots are $n=1,2,3$ spinning BSs. The space covered by NS and BS, although being close and having a border region between them, is different. For similar $\alpha$ (i.e., quadrupolar moments), $\beta$ (the moment of inertia) is always larger for a BS than a NS. Interestingly, there is a region of the plot where the BS branches are closer to the NS ones.  
This corresponds to BS models with large quartic self-interaction, which implies that the corresponding energy-momentum tensor 
approaches a perfect fluid form. Since NS emerge as gravitating solutions where 
matter is given by a perfect fluid energy-momentum tensor, it is not surprising that the two surfaces tend to meet in this limit. We may equivalently say that difference between the BS and NS surfaces could be originated in a non-perfect fluid nature of BS. From the spin parameter point of view, in general, we have higher values for BS. This is by construction; for NS, we can smoothly go from low to high rotation velocities, while for BS this is impossible in our approach.
It is clear that since both NS and BS regions are different, the fitting surfaces allow to break the possible degeneracy between two astrophysical objects through an $I-\chi-Q$ study.

\section{Conclusions}
\label{conclusions}

We investigated the existence of approximate, model-independent, universal, no-hair-like relations between low and high-order multipole moments for various winding number BS. 
First of all, we reaffirmed and extended our previous results
for the $I-\chi-Q$ relations, achieving a better accuracy and confirming the universal behavior with a better than $1.5\%$ precision also for $n=2,3$. We also found effective no-hair relations for the spin octopolar moment, $\chi$ and $Q$ for $n=1,2,3$ with a $3.5\%$, $,1\%$ and $2\%$ precision, respectively, and for the mass hexadecapolar moment,  $\chi$ and $Q$ with a precision of $7\%,1.4\%,1\%$ for $n=1,2,3$. Finally, we found universal behavior in the space $\textit{C}-\chi-Q$ with a precision of $2.5\%,3\%,3\%$, even for stars which are sufficiently compact to possess ergoregions.
We also compared our FIDISOL/CADSOL package calculations with the results of other groups - both those who used the same code and those using other methods - and found agreement whenever a comparison was possible.

For horizonless objects, such universal or effective no-hair relations  allow us to determine the external gravitational field with high precision from a finite number of multipole moments, analogously to the exact no-hair theorems for black holes. This means that even in the presence of matter, it is not necessary to use an infinite number of multipoles to describe the gravitating system within a good approximation.

Right now, the simultaneous measurements of the spin, quadrupole moments, and  moment of inertia, or even the radii of astrophysical objects with high precision, are still difficult challenges.
But keeping the substantial recent progress in GW observations in mind, the universal relations investigated here could play a relevant role
in an astrophysical context in the not-too-distant future.
Independent measurements for two of the quantities linked by any of the relations would allow to obtain the third parameter directly. 
Alternatively, any measurement of all three quantities related by an universal relation would contain valuable information about the nature of the astrophysical object under observation.

An interesting extension of our results would be to study universal relations in the limit in which a horizon has formed inside the rotating BS -- hairy Kerr black holes --  or the investigation of universal relations for the vectorial boson case, the Proca Stars, or some other mixed or exotic compact objects. But the more significant step would be to obtain the tidal and rotational Love numbers within the rapid rotation framework. This is quite interesting from the GW astronomy point of view, but extremely involved since the formalism is based on a perturbation of the full-rotating metric as a base.
As for their NS counterparts, we expect that these universal relations may become helpful in the analysis of gravitational waveforms of future binary merger events, the search for possible bosonic self-coupling terms for dark matter candidates, and in the further understanding of the strong gravity regime of General Relativity.

\begin{acknowledgements}
The authors thank C. Naya for the helpful discussions. JCM thanks Gr@v group at Aveiro University for all the support and hospitality, specially E.Radu, J.Delgado, and E.Costa for their crucial help with the FIDISOL/CADSOL package and further useful comments. JCM also thanks N.Sanchis-Gual for crucial comments and discussions.
Further, the authors acknowledge financial support from the Ministry of Education, Culture, and Sports, Spain (Grant No. PID2020-119632GB-I00), the Xunta de Galicia (Grant No. INCITE09.296.035PR and Centro singular de investigación de Galicia accreditation 2019-2022), the Spanish Consolider-Ingenio 2010 Programme CPAN (CSD2007-00042), and the European Union ERDF.
AW is supported by the Polish National Science Centre,
grant NCN 2020/39/B/ST2/01553.
AGMC is grateful to the Spanish Ministry of Science, Innovation and Universities, and the European Social Fund for funding his predoctoral research activity (\emph{Ayuda para contratos predoctorales para la formaci\'on de doctores} 2019). MHG and JCM thank the Xunta de Galicia (Consellería de Cultura, Educación y Universidad) for funding their predoctoral activity through \emph{Programa de ayudas a la etapa predoctoral} 2021. JCM thanks the IGNITE program of IGFAE for financial support.
\end{acknowledgements}



\appendix
\section{Potentials and numerical parameters used}
\label{appendixA}
In this paper, we have selected a set of physically well-motivated potentials, fitting various astrophysical scenarios, like dark matter haloes, \cite{Mielke:2019rvl,Mielke:2020mve}, BH and NS-like objects \cite{Choi:2019mva,Guerra:2019srj,Delgado:2020udb,Vaglio:2022flq,Grandclement:2014msa}. All of them have been considered in the literature and support different qualitative properties of BS.

\begin{table}[h!]
	\centering
		\begin{tabular}{|c|c|}
			\hline
		 Name & $V\left(\phi\right)$ \\ \hline
			Mini-BS, BS$_{\rm Mass}$& $V_{\rm Mass}=\mu^2\phi^2$ \\ \hline
		BS$_{\rm Quartic}$&$V_{\rm Quartic}=\mu^2\phi^2+|\lambda|/2\phi^4$    \\  \hline
				BS$_{\rm Halo}$& $V_{\rm Halo}=\mu^2\phi^2-|\alpha|\phi^4$  \\ \hline
			BS$_{\rm HKG}$ & $V_{\rm HKG}=\mu^2\phi^2-\alpha\phi^4+\beta\phi^6$  \\ \hline
			BS$_{\rm Sol}$& $V_{\rm Sol}=\mu^2\phi^2(1-(\phi^2/\phi_0^2))^2$  \\ \hline
          BS$_{\rm Sant}$& $V_{\rm Sol}=\mu^2\phi^2(1-(\phi^4/\phi_0^2))^2$  \\ \hline
			BS$_{\rm Log}$&$V_{\rm Log}=f^2\mu^2 \ln\left(\phi^2/f^2+1\right)$\\ \hline
			BS$_{\rm Liouville}$& $V_{\rm Liouville}=f^2\mu^2 \left(\exp\{\phi^2/f^2\}-1\right)$  \\ \hline
				BS$_{\rm Axion}$& $V_{\rm Axion}=\frac{2\mu^2f^2}{B}\left(1-\sqrt{1-4B\sin^2(\phi/2f)}\right)$  \\
			\hline
		\end{tabular}
		\caption{\small BS potentials analyzed in the current work have been previously considered in the case
of spherical, non-rotating BSs, and some cases, further
generalized to rotating solutions   \cite{Siemonsen:2020hcg}. 
Ranging from the so-called \textit{Mini-boson star} potential, through the inclusion of higher order self-interaction terms, e.g. $|\Phi|^4$ and $|\Phi|^6$ \cite{Schunck:2003kk,Colpi:1986ye,Grandclement:2014msa}. Also potentials based on the logarithm, exponential, sine functions and the axion potential\cite{Delgado:2020udb,Choi:2019mva,Guerra:2019srj}. }
		\label{Table.Potentials}
\end{table}

Here we give the different numerical sets of values for the parameters that we have used for our simulations. The numerical values are given in rescaled units.

 \begin{equation}
V_{\rm Quartic}=\phi^2+\frac{\lambda}{2}\phi^4
\begin{cases}
      \lambda= 1\\
      \lambda=10\\
      \lambda=40\\
      \lambda=50\\
      \lambda=60\\
      \lambda=70\\
      \lambda=80
\end{cases}
 \end{equation}

\begin{equation}
V_{\rm Halo}=\phi^2-\alpha\phi^4
\begin{cases}
      \alpha= 1,\\
      \alpha=12.\\
\end{cases}
 \end{equation}

\begin{equation}
V_{\rm HKG}=\phi^2-\alpha\phi^4+\beta\phi^6
\begin{cases}
      \alpha=80, &\beta=0.01\\
      \alpha=2, &\beta=1.8\\
\end{cases}
 \end{equation}

\begin{equation}
V_{\rm Sant}=\phi^2\left(1-\left(\frac{\phi^4}{\phi_0^2}\right)\right)^2
\begin{cases}
      \phi_0=1.5, \\
      \phi_0=0.7,\\
      \phi_0=0.3,\\
       \phi_0=0.1,\\
        \phi_0=0.05.
\end{cases}
 \end{equation}

\begin{equation}
V_{\rm Sol}=\phi^2\left(1-\left(\frac{\phi^2}{\phi_0^2}\right)\right)^2
\begin{cases}
      \phi_0=1.5, 
\end{cases}
 \end{equation}

\begin{equation}
\begin{split}
V_{\rm Axion}&=\frac{2f^2}{B}\left(1-\sqrt{1-4B\sin^2(\frac{\phi}{2f})}\right)\\&
\begin{cases}
      f=0.1, &B=0.22.\\
      f=0.05, &B=0.22.
\end{cases}
\end{split}
 \end{equation}

\begin{equation}
V_{\rm Log}=f^2 \ln\left(\phi^2/f^2+1\right)
\begin{cases}
      f=0.7, \\
      f=0.5.\\
\end{cases}
 \end{equation}

\begin{equation}
V_{\rm Liouville}=f^2 \left(e^\frac{\phi^2}{f^2}-1\right)
\begin{cases}
      f=0.8. \\
\end{cases}
 \end{equation}

\section{Polynomial bases and integrals}
\label{appendixB}
As shown in \cref{multipoles}, each metric function has an expansion in different polynomial bases and coefficients, which can be expanded in radial powers.
When we obtain the $\nu_{2l,k}$ coefficients, we have to integrate $\nu(r,\theta)$ with the corresponding Legendre polynomial and normalization factor. We use the polynomials orthogonality, ensuring that the radial function after the integration, has precisely the correct radial power. After that, we fit with a radial power law extracting the desired coefficient. We show below the orthogonality relations, the normalization factors, and also the first polynomials of each kind:

\textit{Legendre:}
\begin{equation}
    \int_{-1}^{1}P_n(x)P_{m}(x)dx=\frac{2}{2n+1}\delta_{nm}.
\end{equation}
\begin{itemize}
    \item $P_0(x)=1$,\hspace{3.6cm}$N_0=\frac{1}{2}$,
    \item $P_1(x)=x$,\hspace{3.6cm}$N_1=\frac{3}{2}$,
     \item $P_2(x)=\frac{1}{2}(3x^2-1)$,\hspace{2.2cm}$N_2=\frac{5}{2}$,
      \item $P_3(x)=\frac{1}{2}(5x^3-3x)$,\hspace{2cm}$N_3=\frac{7}{2}$,
       \item $P_4(x)=\frac{1}{8}(35x^4-30x^2+3)$,\hspace{1cm}$N_4=\frac{9}{2}$,
\end{itemize}

\textit{Legendre derivatives:}

\begin{equation}
    \int_{-1}^{1}(1-x^2)\frac{dP_n(x)}{dx}\frac{dP_{m}(x)}{dx}dx=\frac{2n(n+1)}{2n+1}\delta_{nm}.
\end{equation}
\begin{itemize}
    \item $\frac{dP_0(x)}{dx}=0$,\hspace{3.6cm}$A_0=0$,
    \item$\frac{dP_1(x)}{dx}=1$,\hspace{3.6cm}$A_1=\frac{3}{4}$,
    \item $\frac{dP_2(x)}{dx}=3x$,\hspace{3.4cm}$A_2=\frac{5}{12}$,
    \item $\frac{dP_3(x)}{dx}=\frac{3}{2}(5x^2-1)$,\hspace{2.2cm}$A_3=\frac{7}{24}$,
    \vspace{0.2cm}
     \item $\frac{dP_4(x)}{dx}=\frac{5}{2}(7x^3-3)$,\hspace{2.2cm}$A_4=\frac{9}{40}$,
   
\end{itemize}

\textit{Gegenbauer polynomials:}

\begin{equation}
    \int_{-1}^{1}(1-x^2)^{\frac{1}{2}}T_n^{\frac{1}{2}}(x)T_m^{\frac{1}{2}}(x)dx=\delta_{nm}.
\end{equation}
And the normalization factors are always $C=1$.
Being:
\begin{equation}
    T_n^{\frac{1}{2}}(x)=\frac{(-1)^n\Gamma(n+2)}{2^{n+\frac{1}{2}}n!\Gamma(n+\frac{3}{2})}(1-x^2)^{-\frac{1}{2}}\frac{d^n}{dx^n}(1-x^2)^{n+\frac{1}{2}}
\end{equation}

\begin{itemize}
    \item $T_0^{\frac{1}{2}}(x)=\sqrt{\frac{2}{\pi}}$
    \item $T_1^{\frac{1}{2}}(x)=2x\sqrt{\frac{2}{\pi}}$,
     \item $T_2^{\frac{1}{2}}(x)=(4x^2-1)\sqrt{\frac{2}{\pi}}$,
      \item $T_3^{\frac{1}{2}}(x)=4x(2x^2-1)\sqrt{\frac{2}{\pi}}$,
       \item $T_4^{\frac{1}{2}}(x)=(16x^4-12x^2+1)\sqrt{\frac{2}{\pi}}$,
\end{itemize}

\section{Fitting coefficients}
\label{appendixC}


\begin{table*}[h!]
	\centering
		\begin{tabular}{|c|c|c|}
			\hline
		 Coeffs & $A_0= 0.1675$ & $B=   3.9567$ \\ \hline
   $A_1^0= -1.6579$	& $A_1^1=  0.0937$  & $A_1^2=  -0.1431$ \\ \hline
	$A_2^0=     -0.7643$& $A_2^1= 0.1706$ &  $A_2^2=   -0.1357 $\\  \hline
	$A_3^0=   0.0382$ & $A_3^1=   -0.0281 $& $A_3^2=  -0.1030$\\ \hline
		\end{tabular}
		\caption{Numerical values of the coefficients that fit the universal $\beta-\chi-\alpha$ surface for $n=1$. }
		\label{Table.Constants12}
\end{table*}

\begin{table*}[h!]
	\centering
		\begin{tabular}{|c|c|c|}
			\hline
		 Coeffs & $A_0= 1.4443$ & $B=   1.7376$ \\ \hline
   $A_1^0= 1.1444$	& $A_1^1= -0.8881$  & $A_1^2=  0.2219$ \\ \hline
	$A_2^0=    1.0449$& $A_2^1= -0.9546$ &  $A_2^2=   0.2547 $\\  \hline
	$A_3^0=   -0.2661$ & $A_3^1=   0.0643 $& $A_3^2=  0.3415$\\ \hline
		\end{tabular}
		\caption{Numerical values of the coefficients that fit the universal $\beta-\chi-\alpha$  surface for $n=2$. }
		\label{Table.Constants11}
\end{table*}

 \begin{table*}[h!]
	\centering
		\begin{tabular}{|c|c|c|}
			\hline
		 Coeffs & $A_0= 1.4724$ & $B=   1.6139$ \\ \hline
   $A_1^0= 1.4724$	& $A_1^1= -1.1080$  & $A_1^2=  0.2458$ \\ \hline
	$A_2^0=     1.6913$& $A_2^1= -1.4560$ &  $A_2^2=   0.3196 $\\  \hline
	$A_3^0=   -0.5509$ & $A_3^1=   0.1169 $& $A_3^2=  0.6692$\\ \hline
		\end{tabular}
		\caption{Numerical values of the coefficients that fit the universal $\beta-\chi-\alpha$  surface for $n=3$. }
		\label{Table.Constants10}
\end{table*}

\begin{table*}[h!]
	\centering
		\begin{tabular}{|c|c|c|}
			\hline
		 Coeffs & $A_0= 1.1182$ & $B=   -6.8583$ \\ \hline
   $A_1^0= -1.0416$	& $A_1^1=  1.6527$  & $A_1^2=-0.7404$ \\ \hline
	$A_2^0=  3.2635$& $A_2^1= -5.6762$ &  $A_2^2=2.5947 $\\  \hline
	$A_3^0= -4.2039$ & $A_3^1=   7.5973 $& $A_3^2=  -3.4679$\\ \hline
      $A_4^0= 1.5036$ & $A_4^1= -3.0087 $& $A_4^2=  1.4358$\\ \hline
		\end{tabular}
		\caption{Numerical values of the coefficients that fit the universal  $\sqrt[3]{\Bar{s}_3}-\chi-\Bar{Q}$  surface for $n=1$.}
		\label{Table.Constants9}
\end{table*}

\begin{table*}[h!]
	\centering
		\begin{tabular}{|c|c|c|}
			\hline
		 Coeffs & $A_0= 1.0370$ & $B=  3.3126$ \\ \hline
   $A_1^0= -0.3622$	& $A_1^1=0.8047$  & $A_1^2=-0.4139$ \\ \hline
	$A_2^0=  -5.0622$& $A_2^1=3.4735$ &  $A_2^2=0.2927 $\\  \hline
	$A_3^0= 47.3622$ & $A_3^1=-51.4001 $& $A_3^2=12.1754 $\\ \hline
        $A_4^0= -89.6563$ & $A_4^1=103.1442 $& $A_4^2=-28.1142$\\ \hline
		\end{tabular}
		\caption{Numerical values of the coefficients that fit the universal  $\sqrt[3]{\Bar{s}_3}-\chi-\Bar{Q}$  surface for $n=2$.}
		\label{Table.Constants8}
\end{table*}

 \begin{table*}[h!]
	\centering
		\begin{tabular}{|c|c|c|}
			\hline
		 Coeffs & $A_0= 1.0462$ & $B=   2.5451$ \\ \hline
   $A_1^0=-0.4942$	& $A_1^1= 1.1952$  & $A_1^2= -0.6359$ \\ \hline
	$A_2^0=-3.5220$& $A_2^1= -1.237$ &  $A_2^2=  3.5191 $\\  \hline
	$A_3^0=  41.0614 $ & $A_3^1= -44.7257 $& $A_3^2= 4.2756$\\ \hline
       $A_4^0=  13.7087 $ & $A_4^1= 23.0641 $& $A_4^2= -5.2694$\\ \hline
		\end{tabular}
		\caption{Numerical values of the coefficients that fit the universal  $\sqrt[3]{\Bar{s}_3}-\chi-\Bar{Q}$  surface for $n=3$.}
		\label{Table.Constants7}
\end{table*}

\begin{table*}[h!]
	\centering
		\begin{tabular}{|c|c|c|c|}
			\hline
		 Coeffs & $A_0= -1.3490$ & $B=   -9.6286$ &$\sim$\\ \hline
   $A_1^0=-0.3302$	& $A_1^1= 2.0855$  & $A_1^2= -2.0064$  & $A_1^3= 0.5953$ \\ \hline
	$A_2^0= 1.1406$& $A_2^1=-9.1543$ &  $A_2^2= 9.3758 $ &  $A_2^3=-2.6856 $ \\  \hline
	$A_3^0= 0.9853$ & $A_3^1= 12.6495 $& $A_3^2=-16.6059$  &  $A_3^3= 5.2729 $  \\ \hline
       $A_4^0= -7.7049$ & $A_4^1= 6.5678 $& $A_4^2= 0.9884$  &  $A_4^3=  -1.3193$ \\ \hline
		\end{tabular}
		\caption{Numerical values of the coefficients that fit the universal  $\sqrt[4]{\Bar{m}_4}-\chi-\Bar{Q}$  surface for $n=1$. }
		\label{Table.Constants6}
\end{table*}

\begin{table*}[h!]
	\centering
         \resizebox{8.5cm}{!} {
		\begin{tabular}{|c|c|c|c|}
			\hline
		 Coeffs & $A_0=1.1219$ & $B=  1.4261$ &$\sim$\\ \hline
   $A_1^0= -0.5964$	& $A_1^1= 2.3931$  & $A_1^2=  -2.3414$  & $A_1^3= 0.7396$ \\ \hline
	$A_2^0= -8.0427$& $A_2^1= 7.8986$ &  $A_2^2=-0.9397 $ &  $A_2^3= -1.0319 $ \\  \hline
	$A_3^0=  72.8292$ & $A_3^1= -104.9515$& $A_3^2=53.0559$  &  $A_3^3= -8.3977$  \\ \hline
       $A_4^0=  -225.0507$ & $A_4^1=349.2477 $& $A_4^2=-187.1246$  &  $A_4^3=  33.8346$ \\ \hline
		\end{tabular}}
		\caption{Numerical values of the coefficients that fit the universal  $\sqrt[4]{\Bar{m}_4}-\chi-\Bar{Q}$  surface for $n=2$.  }
		\label{Table.Constants5}
\end{table*}

 \begin{table*}[h!]
	\centering
 \resizebox{8.5cm}{!} {
		\begin{tabular}{|c|c|c|c|}
			\hline
		 Coeffs & $A_0= 1.1910$ & $B=   1.7051$ &$\sim$\\ \hline
   $A_1^0=-0.7087$	& $A_1^1=  2.8764$  & $A_1^2= -2.9818$  & $A_1^3=  1.0053$ \\ \hline
	$A_2^0=  -11.7062  $& $A_2^1= 15.2705$ &  $A_2^2= -4.2939 $ &  $A_2^3=  -1.4674 $ \\  \hline
	$A_3^0=  42.4967 $ & $A_3^1=  -58.0106 $& $A_3^2= 31.9637$  &  $A_3^3= -2.9678$  \\ \hline
       $A_4^0= -110.2867$ & $A_4^1= 115.9387$& $A_4^2= -52.5249$  &  $A_4^3=  7.1676$ \\ \hline
		\end{tabular}}
		\caption{Numerical values of the coefficients that fit the universal $\sqrt[4]{\Bar{m}_4}-\chi-\Bar{Q}$  surface for $n=3$.  }
		\label{Table.Constants4}
\end{table*}

\begin{table*}[h!]
	\centering
		\begin{tabular}{|c|c|c|c|}
			\hline
		 Coeffs & $A_0= 0.1821$ & $B=  1.8200$ &$\sim$\\ \hline
   $A_1^0= -0.5127$	& $A_1^1=  0.7567$  & $A_1^2= 0.1997$  & $A_1^3=   -51.0979$ \\ \hline
	$A_2^0=    -1.9371$& $A_2^1= 1.5161$ &  $A_2^2=   14.0300 $&  $A_2^3=  -95.5285 $\\  \hline
	$A_3^0=   -0.4379$ & $A_3^1=   3.0799 $& $A_3^2=  24.3730$ & $A_3^3=  -56.9695$\\ \hline
        $A_4^0=   -0.1332$ & $A_4^1=   0.8977 $& $A_4^2=  8.8889$ & $A_4^3=  -10.9534$\\ \hline
		\end{tabular}
		\caption{Numerical values of the coefficients that fit the universal $\sqrt{\textit{C}}-\log\chi-\log\Bar{Q}$  surface for $n=1$.}
		\label{Table.Constants3}
\end{table*}

\begin{table*}[h!]
	\centering
		\begin{tabular}{|c|c|c|c|}
			\hline
		 Coeffs & $A_0=0.2068$ & $B=   1.5938$ &$\sim$\\ \hline
   $A_1^0= -1.8892$	& $A_1^1=  -9.4814$  & $A_1^2=  1.6036$  & $A_1^3=  -61.6216$ \\ \hline
	$A_2^0=    3.2616$& $A_2^1= -3.2824$ &  $A_2^2=   32.8531 $&  $A_2^3=   18.6472 $\\  \hline
	$A_3^0=   -0.0486$ & $A_3^1=   -7.1066 $& $A_3^2=  0.5532$ & $A_3^3=  88.3945$\\ \hline
        $A_4^0=   -0.0485$ & $A_4^1=   -2.1083 $& $A_4^2=  -5.6110$ & $A_4^3=  45.5493$\\ \hline
		\end{tabular}
		\caption{Numerical values of the coefficients that fit the universal  $\sqrt{\textit{C}}-\log\chi-\log\Bar{Q}$  surface for $n=2$. }
		\label{Table.Constants2}
\end{table*}

 \begin{table*}[h!]
	\centering
		\begin{tabular}{|c|c|c|c|}
			\hline
		 Coeffs & $A_0= 0.1910$ & $B=   1.6136$ &$\sim$\\ \hline
   $A_1^0= -1.4963$	& $A_1^1=  -8.7667$  & $A_1^2=  1.2767$  & $A_1^3=  -47.0880$ \\ \hline
	$A_2^0=    4.2726$& $A_2^1=-4.0039$ &  $A_2^2=   39.0807 $&  $A_2^3=   -38.5053 $\\  \hline
	$A_3^0=   0.2126$ & $A_3^1=   -8.1408 $& $A_3^2=  18.2189$ & $A_3^3=  -39.3063$\\ \hline
        $A_4^0=   0.0163$ & $A_4^1=   -2.2857$& $A_4^2=  3.6033$ & $A_4^3=  -12.9732$\\ \hline
		\end{tabular}
		\caption{Numerical values of the coefficients that fit the universal  $\sqrt{\textit{C}}-\log\chi-\log\Bar{Q}$  surface for $n=3$.}
		\label{Table.Constants1}
\end{table*}


\vspace{5cm}


%

\end{document}